\documentclass[runningheads]{llncs}

\usepackage[T1]{fontenc}
\usepackage{graphicx}

\usepackage{textcomp}
\usepackage{xcolor}
\usepackage{appendix}

\usepackage{newtxtext,newtxmath}
\usepackage{subcaption}
\usepackage{comment}

\usepackage{stmaryrd}
\usepackage{bm}

\usepackage{siunitx}
\usepackage{comment}
\usepackage{booktabs}
\usepackage{cite}

\usepackage{xtab,booktabs}

\usepackage[font=footnotesize]{caption} 

\usepackage{placeins}
\usepackage{dblfloatfix}
\usepackage[many]{tcolorbox}
\usepackage{dashrule}

\usepackage{anyfontsize}
\usepackage{bmpsize}
\usepackage{url}
\usepackage{multirow}
\usepackage{enumerate}
\usepackage{enumitem}
\usepackage{listings}
\usepackage{here}
\usepackage{autobreak}
\usepackage{array}
\usepackage{latexsym}
\usepackage{newtxtext}
\usepackage{diagbox}
\usepackage{algpseudocode}
\usepackage{lscape}
\usepackage{tcolorbox}
\usepackage{multicol}
\usepackage{tabularx}
\usepackage{float}
\usepackage{bm} 
\usepackage[skip=2pt]{caption}
\usepackage{makecell}
\usepackage{wrapfig}

\def\BibTeX{{\rm B\kern-.05em{\sc i\kern-.025em b}\kern-.08em
    T\kern-.1667em\lower.7ex\hbox{E}\kern-.125emX}}

\begin{document}
\title{
LLMs, You Can Evaluate It! 
Design of Multi-perspective Evaluation for Reports in Security Operation Centers
}
\titlerunning{LLMs, You Can Evaluate It! }
%
\author{Hiroyuki Okada \and
Tatsumi Oba \and
Naoto Yanai}
\authorrunning{H. Okada et al.}
%
\institute{Panasonic Holdings Corporation, Osaka, Japan \\
\email{okada.hiroyuki001@jp.panasonic.com}
}
\maketitle              
\begin{abstract}
Security operation centers (SOCs) often produce analysis reports on security alerts, and large language models (LLMs) will likely be used for this task. 
While producing analysis reports with actionable insights requires constantly evaluating the reports of SOC practitioners, it is unclear whether LLMs appropriately evaluate the reports due to a lack of evaluation criteria. 
In this paper, we discuss evaluations using LLMs for analysis reports. 
To this end, we first design \textit{Analyst-wise Checklist} that contains SOC practitioners' evaluation criteria for analysis reports in the real world through literature review and interviews with SOC practitioners as a user study. 
Next, we design a novel LLM-based multi-perspective evaluation framework, named \textit{MESSALA}, that maximizes report evaluation and provides feedback with analysis reports on veteran SOC practitioners' perceptions by leveraging the checklist above. 
We also conduct quantitative evaluation and qualitative evaluation with real-world analysis reports. 
For the quantitative evaluation, we demonstrate that MESSALA can appropriately rate each analysis report on the evaluation criteria from the perspective of veteran SOC practitioners compared with existing LLM-based methods. 
For the qualitative evaluation, we identify that MESSALA can provide feedback comments that are actionable for improving analysis reports under the evaluation criteria. 


\end{abstract}

\section{Introduction}
\label{sec:introduction}


Cyberattacks have increased across many organizations, and establishing a security operation center (SOC) to analyze security alerts and provide their responses has become urgent and crucial for each organization in recent years.
An important mission for SOCs is, in addition to providing security alerts and their responses, to write analysis reports with clear and actionable insights into their underlying cyberattacks~\cite{sharma2024decoding}. 
Interestingly, in proportion to the quality of analysis reports, various SOC practitioners, including not only analysts but also their managers, can understand the content of security alerts, and hence, the entire organization will be aware of cyberattacks~\cite{jawad2024m}.

However, writing analysis reports for security alerts forces a heavy workload on SOC practitioners, because these reports must contain actionable insights~\cite{kokulu2019MatchedandMismatchedSOCs}. 
To reduce this workload, tools, such as large language models (LLMs), that introduce actionable insights into analysis reports are in high demand~\cite{albanese2025towardsAI-driven}. 
Although there are several tools to generate analysis reports for security alerts~\cite{loumachi2025advancing,michelet2024chatgpt}, automatically generated reports often contain inaccurate information, which may degrade performance in SOCs~\cite{kokulu2019MatchedandMismatchedSOCs}. 
Even when state-of-the-art LLMs are used, they often cause hallucinations, such as inaccurate information~\cite{huang2025hallucination}. 
As another motivation, writing analysis reports to the exact requirements of report-writing guidelines is stressful~\cite{nepal2024burnout}. 
The reason is that evaluation of analysis reports should identify any lack of descriptions as well as judgments from the perspectives of veteran SOC practitioners~\cite{RYAN2012Quantifying}. 
Although the most common mitigation approach for the above stress is to constantly evaluate analysis reports~\cite{kokulu2019MatchedandMismatchedSOCs}, evaluations of the reports need evaluation criteria based on expert knowledge~\cite{kerstenfield2025tier1analyst,kersten2023give}. 
Moreover, the current guidelines are often abstract and may fail to capture the evaluation criteria from the perspective of veteran SOC practitioners~\cite{johnson2016guide,agyepong2020towards}. 
Namely, even evaluations of analysis reports using LLMs are challenging due to a lack of enough evaluation criteria. 

Based on the above background, in this paper, we discuss evaluations for analysis reports in SOCs using LLMs in order to improve the quality of the reports. 
Specifically, we discuss the following research questions in this paper: 



\begingroup
\begin{description}
\item[RQ1] What are the evaluation criteria that guarantee the quality of analysis reports from SOC practitioners’ knowledge?
\item[RQ2] Can LLMs quantitatively evaluate the quality of analysis reports from the perspective of SOC practitioners under the defined evaluation criteria?
\item[RQ3] Can LLMs qualitatively evaluate the quality of analysis reports with feedback from the perspective of SOC practitioners under the defined evaluation criteria?
\end{description}
\endgroup

We first design the \textit{Analyst-wise Checklist}, which contains the evaluation criteria in its specific items to reflect SOC practitioners’ knowledge for evaluating analysis reports in the real world, as the answer to RQ1. 
To design this, we conducted a literature review of 13 public guidelines and handbooks in the IT and OT domains and semi-structured interviews with 15 SOC practitioners. 
We also conducted a validation test to confirm the checklist with multiple SOC practitioners. 
(See Section~\ref{sec:EvaluationwithAnalyst-Wise} for details.)



Next, based on the above checklist, we propose a novel framework, called \textit{Multi-perspective Evaluation System for Security Analysis using Llm Assistance (MESSALA)}, to maximize the evaluation of analysis reports in SOCs. 
In a nutshell, MESSALA guides LLMs in evaluating analysis reports by leveraging the Analyst-wise Checklist from multiple perspectives. 
As described in Section~\ref{sec:proposed}, MESSALA can imitate SOC practitioners' cognition~\cite{zhong2016automate} to obtain expert knowledge from the checklist, and therefore, can return appropriate evaluation results with actionable feedback for each analysis report. 
When we conduct quantitative evaluations with MESSALA, we show that it can rate analysis reports for the evaluation criteria in comparison with the existing methods~\cite{fu2023gptscore,liu2023g-eval}. 
Thus, to answer RQ2, we confirm if MESSALA enables LLMs to quantitatively evaluate analysis reports from the perspective of veteran SOC practitioners. 
(See Section~\ref{sec:quantitativeevaluation} for details.)

Third, we conduct qualitative evaluations to identify whether the feedback generated by MESSALA aligns with judgments from the perspective of veteran SOC practitioners. 
As a result, we confirm if MESSALA can provide more actionable feedback to both novice and veteran SOC practitioners than the existing methods, as it is more specific and easier to understand. 
It means that MESSALA also enables LLMs to qualitatively evaluate analysis reports through its feedback. 
(See Section~\ref{sec:qualitativeevaluation} for details.)

\color{black}

To sum up, we make the following contributions in this paper: 
\begin{itemize}
    \item We design the Analyst-wise Checklist, which contains evaluation criteria for evaluating analysis reports in order to reflect real-world SOC practitioners' knowledge through literature review and interviews with SOC practitioners, as the answer to RQ1. 
    
    \item We propose MESSALA, a novel framework, that maximizes evaluation for each criterion and provides feedback from the perspective of SOC practitioners, and then confirm if MESSALA enables LLMs to quantitatively evaluate analysis reports under the evaluation criteria.
    This is identical to the answer to RQ2. 
    


    \item We qualitatively evaluate the feedback returned from MESSALA, and then identify that MESSALA enables LLMs to qualitatively evaluate analysis reports through its feedback under the evaluation criteria, which is the answer to RQ3. 
\end{itemize}


\section{Related Work and Our Problem Setting}
\label{sec:LLM-BasedAnalysis}

In this section, we provide background on related work on LLMs and SOCs. 
We then describe the evaluation of analysis reports in SOCs as the main problem setting. 


\subsubsection{Large Language Models}
LLMs generate text outputs from natural-language prompts through probability distributions of tokens.
Various use cases have been proposed; for example, LLMs provide performance comparable to that of veteran data analysts~\cite{cheng2023-IsGPTaGoodDataAnalyst} and can be used as LLM-as-a-judge services for data evaluation~\cite{kim2023prometheus}.
We focus on GPTScore~\cite{fu2023gptscore} and G-Eval~\cite{liu2023g-eval}. 
A common problem for LLMs is hallucination~\cite {huang2025hallucination}, in which LLMs generate outputs that are inadequate or incorrect. 

LLMs have been utilized in many cybersecurity research, such as 
data collection~\cite{BAYER2023multi-levelfine-tuning,shibli2024AbuseGPT}, 
attack and threat analysis tasks~\cite{gadyatskaya2023ChatGPTKnowsYourAttacks,BOFFA2024Logprecis,shibli2024AbuseGPT},
and report generation tasks~\cite{Gupta2023fromChatGPTtoThreatGPT,mitra2024LocalIntel,perrina2023agir,loumachi2025advancing,wickramasekara2024framework}. 
However, the above works do not provide a detailed discussion of evaluations of analysis reports.
We focus on the evaluation of reports in this paper.

\subsubsection{Security Operation Centers}


SOCs address various aspects of cyberattacks, including data collection, determining the presence of intrusions, and responding to threats~\cite{Tilbury2024HumansandAutomation}.
The current SOCs monitor industrial control systems, such as factories, as well as conventional IT systems~\cite{albasheer2022Cyber-attack,oba2024ScoreAndYouShallFind}. 
The major role of SOCs is to detect cyberattacks~\cite{gupta2019AutomatedEventPrioritization,renners2019AdaptiveandIntelligible}, investigate security alerts~\cite{kerstenfield2025tier1analyst,yen2013Beehive}
, and make judgments on these alerts~\cite{kokulu2019MatchedandMismatchedSOCs}. 
SOCs roughly consist of two processes, i.e., the evaluation process by analysts and the judgment process by managers. 
To support and connect both processes, analysis reports are significantly important~\cite{kokulu2019MatchedandMismatchedSOCs}.

The major challenges of SOCs are to detect novel cyberattacks~\cite{gupta2019AutomatedEventPrioritization,renners2019AdaptiveandIntelligible} and reduce the volume of security alerts that analysts must triage~\cite{kokulu2019MatchedandMismatchedSOCs,yen2013Beehive,zhong2019LearningfromExpertsExperimence}. 
Although SOCs in recent years have introduced automation tools~\cite{singh2025contextbuddy,turcotte2025automated,shahjee2022IntegratedNetwork,AGYEPONG2023ASystematicMethod} to reduce workload~\cite{Tilbury2024HumansandAutomation} for response to security alerts, it is still heavy~\cite{nepal2024burnout,kerstenfield2025tier1analyst,kersten2024securityalert}.
Analysts in SOCs also need to understand cyberattacks based on fragmented information~\cite{VANDERKLEIJ2022DevelopingDecisionSupport} and frequently require source information to determine appropriate responses~\cite{happa2021Assessing}. 
Although machine learning, such as LLMs, has been developed for analysis in SOCs recently~\cite{ANDRADE2019CognitiveSecurity,gonzalez2021SIEM,HUSAK2022Crusoe,mitra2024LocalIntel,albanese2025towardsAI-driven,singh2025contextbuddy}, 
there are limitations for providing analysis reports~\cite{michelet2024chatgpt,loumachi2025advancing,wickramasekara2024framework} and security advice~\cite{chen2023can} from LLMs due to several reasons, such as hallucinations described above. 

\subsubsection{Evaluation of Analysis Reports in SOCs}

We focus on evaluations of analysis reports in SOCs as the main problem setting. 
The current report generation tools~\cite{loumachi2025advancing,michelet2024chatgpt,wickramasekara2024framework} have limitations, because LLMs often cause hallucinations and these tools also lack explicit mechanisms for evaluating of analysis reports. 
Instead, we aim to explore whether LLMs can appropriately evaluate these reports.
We note that evaluating analysis reports remains challenging: general LLMs may fail to align with human judgment because these reports include domain-specific context~\cite{chiang2023closer}.

\paragraph{Problem Setting}
We aim to design an evaluation framework that takes analysis reports as input and returns evaluation results as output. The output includes both quantitative scores and qualitative feedback from the perspectives of SOC practitioners, including veteran analysts. 
We consider that readers of analysis reports are both analysts and managers, while the generation of the reports is outside the scope of this paper.

\begingroup

\section{Designing Analyst-wise Checklist for Security Report Evaluation}
\label{sec:EvaluationwithAnalyst-Wise}

This section aims to clarify the criteria for evaluating the quality of security alert analysis reports and to construct the Analyst-wise Checklist that reflects these criteria. 
The following section constitutes the Response to RQ1.
This type of report is a highly specialized document that presupposes advanced knowledge of the security domain. 
It has been noted that general text evaluation metrics are often insufficient for evaluating the quality of domain-specific reports~\cite{stufflebeam2000guidelines}.
The Analyst-wise Checklist contains practical evaluation criteria derived from the literature review and interviews and is also effective for decision-making by stakeholders.

\subsection{Design Process of the Analyst-wise Checklist}
The Analyst-wise Checklist is constructed by the following four processes. 
This approach is a common practice for incorporating expert knowledge and domain-specific context into design~\cite{benton2020usability,kwong2025design,rose2022co}. 
1) Collection of Candidate Items: We first collected literature in related research fields and then formulated both the information that should be documented in analysis reports and an initial set of evaluation items.
2) Interviews on Report Quality: Semi-structured interviews were conducted with SOC practitioners to extract practical and critical evaluation criteria with respect to the quality of analysis reports.
3) Development of the Analyst-wise Checklist Draft:
Based on the evaluation criteria extracted from the above processes, we newly designed concrete checklist items that are particularly important for evaluating analysis report quality.
4) Refinement through Expert Review and Evaluation Tests:
For the checklist draft, we revised its item structure and wording based on expert review, and then conducted trial evaluations in which multiple authors independently applied the checklist to report samples. 
After refining problematic items, we finalized the checklist items through agreement between the authors and experts. 
The details of these processes are described below.




\subsection{Collection of Candidate Items through Literature Review}
The initial process of designing the checklist involves a literature review. 
This literature review aims to establish the design foundation of the checklist, and the process consists of the following steps:

1) Literature Collection: To identify the knowledge required for the checklist items, we collected official guidelines for incident response and SOC operations, which were released by public organizations and expert communities. 
They include NIST SP 800-series incident handling guidelines and public information on incident response from domestic and international CSIRTs, regulatory bodies, and industry groups~\cite{meti_is_standards_2016,johnson2003handbook,johnson2016guide,scarfone2008sp,cisa_incident_reporting_2024,enisa_eecc_incident_2021,uscert_incident_guidelines_2015,enisa_incident_management_2010,isogj_soc_textbook_2023,isogj_5w1h_2019,mitre_11strategies_2022,cisa_playbooks_2021,first_csirt_framework_2019}.

First, We collected 35 public guidelines and incident handbooks
in 2000-2024 from government agencies, leading authorities, and expert communities. These documents were sourced from their official websites
and technical repositories. 
After a screening process based on keyword searches described in Item 2), we selected thirteen representative guidelines.

2) Extraction of Relevant Content: To efficiently extract important checklist items, we combined keyword searching using several words, such as ``report," ``should be described," ``should be included," and ``information sharing," with manual reading by the authors. 
The above step extracts sentences for the information that should be documented in analysis reports. 
We also use an LLM to follow several guidelines with long pages: specifically, we query the LLM with the prompt as "List the information that should be included in the incident report envisioned by this guideline, and specify where it is written." 
This provides a candidate set of the checklist items.

3) Manual Selection of Statements: To ensure clear and concrete checklist items, the sentences obtained from the keyword searching and the candidate set of the checklist items are manually checked by the authors themselves. 
By cross-referencing them with the original texts in each guideline, we selected only the descriptions corresponding to information that should be documented in analysis reports.

4) Organization of Information: The descriptions obtained in the previous step were systematically coded by two of the authors with respect to the items that should be documented in an analysis report. 
Through this coding process, the extracted descriptions were conceptually grouped according to their functional role within the analysis reports and organized into three categories and eleven concrete items. 
These three categories and eleven items were reviewed and revised by two SOC practitioners. 
Afterward, we further collected several guidelines to expand knowledge from the initial collection of literature. 
Since this expanded step did not yield any new content on the above three categories and eleven items, we regarded the above categories and items as complete with respect to the evaluation criteria.
We present these three categories and eleven items that should be documented in analysis reports below.

\textbf{Decision support and action planning:}
This category groups report elements that help frontline staff and managers grasp the current situation and determine appropriate next steps. 
The checklist items assess whether recommended actions and follow-up are concrete and well-justified, and whether they enable immediate response with long-term improvement.
In particular, the checklist items include Analysis Status\cite{meti_is_standards_2016,johnson2003handbook,scarfone2008sp,uscert_incident_guidelines_2015,mitre_11strategies_2022,cisa_playbooks_2021,first_csirt_framework_2019}, Impact Assessment~\cite{johnson2003handbook,johnson2016guide,scarfone2008sp,cisa_incident_reporting_2024,enisa_eecc_incident_2021,uscert_incident_guidelines_2015,enisa_incident_management_2010,isogj_soc_textbook_2023,isogj_5w1h_2019,mitre_11strategies_2022,cisa_playbooks_2021,first_csirt_framework_2019}, Confirmation Requests~\cite{cisa_incident_reporting_2024,enisa_incident_management_2010,isogj_5w1h_2019,first_csirt_framework_2019}, Response Actions~\cite{meti_is_standards_2016,johnson2003handbook,scarfone2008sp,cisa_incident_reporting_2024,uscert_incident_guidelines_2015,enisa_incident_management_2010,isogj_soc_textbook_2023,isogj_5w1h_2019,mitre_11strategies_2022,cisa_playbooks_2021,first_csirt_framework_2019}, and Recurrence Prevention \& Lessons~\cite{meti_is_standards_2016,johnson2003handbook,johnson2016guide,scarfone2008sp,cisa_incident_reporting_2024,enisa_eecc_incident_2021,enisa_incident_management_2010,isogj_soc_textbook_2023,isogj_5w1h_2019,mitre_11strategies_2022,cisa_playbooks_2021,first_csirt_framework_2019}.

\textbf{Accountability and quality assurance of the investigation and analysis process:}
This category ensures that investigation activities and analysis judgments are documented in a transparent and traceable manner for SOC operations. 
The checklist items assess whether the process is described systematically and with sufficient detail to enable third-party review, compliance verification, and future incident handling.
In particular, the checklist items include Investigation \& Analysis Methods~\cite{johnson2003handbook,scarfone2008sp,enisa_incident_management_2010,isogj_soc_textbook_2023,isogj_5w1h_2019,mitre_11strategies_2022,cisa_playbooks_2021,first_csirt_framework_2019}, Evidence \& Supporting Data~\cite{johnson2003handbook,cisa_incident_reporting_2024,enisa_eecc_incident_2021,uscert_incident_guidelines_2015,isogj_soc_textbook_2023,isogj_5w1h_2019}, Related Policies~\cite{scarfone2008sp,enisa_eecc_incident_2021,isogj_soc_textbook_2023,isogj_5w1h_2019,cisa_playbooks_2021,first_csirt_framework_2019}, and Guidelines \& Standards~\cite{johnson2003handbook,cisa_incident_reporting_2024,enisa_incident_management_2010,isogj_soc_textbook_2023,first_csirt_framework_2019}. 

\textbf{Technical understanding and root cause clarification of the event:}
This category covers the technical description of what happened, where, to whom, and how. 
The checklist items assess whether the incident is described concretely enough to allow technical reconstruction and to justify the reported conclusions.
In particular, the checklist items include Event Description \& Interpretation~\cite{johnson2003handbook,johnson2016guide,scarfone2008sp,cisa_incident_reporting_2024,enisa_eecc_incident_2021,uscert_incident_guidelines_2015,isogj_soc_textbook_2023,isogj_5w1h_2019,cisa_playbooks_2021,first_csirt_framework_2019}, Root Cause~\cite{johnson2003handbook,scarfone2008sp,cisa_incident_reporting_2024,enisa_eecc_incident_2021,isogj_soc_textbook_2023,cisa_playbooks_2021,first_csirt_framework_2019,enisa_incident_management_2010}, Incident Source \& Impacted Systems~\cite{johnson2003handbook,scarfone2008sp,enisa_eecc_incident_2021,uscert_incident_guidelines_2015,isogj_soc_textbook_2023,isogj_5w1h_2019,cisa_playbooks_2021,first_csirt_framework_2019,enisa_incident_management_2010}, Communication Details\cite{johnson2016guide,scarfone2008sp,isogj_5w1h_2019}, and Vulnerability Information~\cite{johnson2016guide,scarfone2008sp,enisa_eecc_incident_2021,enisa_incident_management_2010,isogj_soc_textbook_2023,isogj_5w1h_2019,mitre_11strategies_2022,cisa_playbooks_2021,first_csirt_framework_2019}
.

\subsection{Interviews on Report Quality}

\begin{table*}[t]
    \caption{Participant information.
    For the row of Expertise, we categorize participants as ``High", ``Middle", and ``Low" based on their self-reported expertise and years of experience.
    }
    \label{tab:userprofile}
    \centering
    \begin{tabular}{llll}
        \hline
        \makecell[l]{\textbf{User} \\ \textbf{ID}} &
        \makecell[l]{\textbf{Expertise} \\ \textbf{/ Experience(years)}} &
        \makecell[l]{\textbf{Job Title}} &
        \makecell[l]{\textbf{Target Environment}} \\ \hline
        1  & High / 7     & SOC Manager, Architect                & Factory, Building \\
        2  & High / 5   & Engineer, SOC Analyst                 & Building, Home Appliances \\
        3  & High / 9     & SOC Manager, Architect                & IT, Factory \\
        4  & High / 6   & SOC Analyst, Supervisor               & IT \\
        5  & Middle / 4 & Engineer, SOC Analyst                 & Factory, Home Appliances \\
        6  & High / 17     & SOC Analyst                           & Data Center \\
        7  & High / 23     & SOC Analyst                           & Data Center \\
        8  & Low / 3       & SOC Analyst                           & IT, Factory \\
        9  & High / 5      & SOC Analyst, Supervisor               & IT, Factory \\
        10 & High / 12     & SOC Analyst, Supervisor               & IT \\
        11 & High / 20     & FSIRT/CSIRT Staff                     & Factory \\
        12 & Middle / 2    & PSIRT Staff                           & Transportation, Cold Chain \\
        13 & High / 20   & Control Vendor Security Staff         & Building \\
        14 & Low / 3    & Security Manager (Manufacturing Site) & IT, Factory \\
        15 & Low / 2    & SOC Analyst, Network Engineer          & IT, Factory \\
        \hline
    \end{tabular}
\end{table*}

We conducted semi-structured interviews with SOC practitioners responsible for drafting and reviewing analysis reports in order to develop the Analyst-wise Checklist. 
It can reflect SOC practitioners’ concerns about analysis reports and the operational context into the checklist~\cite{rose2022co}. 
More specifically, the purpose of these interviews is to elicit the implicit evaluation criteria that SOC practitioners use when judging the quality of analysis reports.
One might think that generic criteria and metrics for text quality can be applied to the three categories and eleven items obtained in the previous process. 
However, we focus on the perspectives that SOC practitioners strongly recognize as problematic in their work, and prioritize clarifying which aspects should be checked most carefully. 
For example, while purely formal aspects such as spelling and typographical errors are reasonably important, it is more likely that the quality of analysis reports is determined by content-related factors, such as adequate descriptions of the operational context and feasible actions. 
It is therefore essential to capture these evaluation criteria appropriately in the checklist items.

\subsubsection{Study Design}
We conducted interviews with fifteen participants from eight different SOC-related organizations. 
Participants included veteran analysts, managers, and SIRT members, all of whom were involved in at least one stage of analysis reports, such as authoring, reviewing, and approving reports. 
The participants were recruited via an online interview platform from IT/OT/IoT environments, including the authors’ affiliations and partner companies. 
Detailed organizational attributes of the participants are shown in Table~\ref{tab:userprofile}.
The interviews were conducted by one or more authors in order to mitigate potential biases in the questions and their interpretation. Participants were recruited using online interview platforms, such as SPRINT and unii research, and the interviews themselves were conducted using communication tools including Microsoft Teams.
Upon the participants’ consent, all the interviews were audio-recorded and subsequently transcribed, yielding a total of 337 comments. 
Prior to analysis, the transcripts were anonymized and stored with appropriate security safeguards for research ethics. 
For the participants from organizations outside the authors’ affiliations, we provided a reward of a common amount for all the participants. 
To encourage discussion based on the conversation of interviews, we adopted a semi-structured interview format with approximately thirty to sixty minutes. 
An actual template of interview questions are shown in Table~\ref{tab:questionsheet}.
\begin{table*}[t!]
    \caption{Question Sheet for User Background and Preliminary User Study}
    \label{tab:questionsheet}
    \footnotesize
    \centering
    \begin{tabularx}{\textwidth}{|
        p{2cm}  
        |p{1cm} 
        |X|}
        \hline
        \textbf{Category} & \textbf{No.} & \textbf{Summary} \\
        \hline

        \multirow{10}{=}{User Information}
        & 1-1. & How many years of experience do you have in security alert analysis? \\
        \cline{2-3}
        & 1-2. & What areas of security monitoring do you do? (IT (various industries), IoT (factories, electric power, cars, etc.)) \\
        \cline{2-3}
        & 1-3. & What is the role of the SOC? \\
        \cline{2-3}
        & 1-4. & What is your specialty? (Cybersecurity, Software Engineering, Computer Science, etc.) \\
        \cline{2-3}
        & 1-5. & What types of security alerts do you primarily cover? \\
        \cline{2-3}
        & 1-6. & What types of reports do you provide to your customers? \\
        \cline{2-3}
        & 1-7. & How often do you report to your customers? \\
        \cline{2-3}
        & 1-8. & What is the process for creating a security alert analysis report? \\
        \cline{2-3}
        & 1-9. & Where do you get the information you need to create reports? (e.g., SIEMs, log management systems, manual investigations) \\
        \cline{2-3}
        & 1-10. & How long does it take to create a single report? \\
        \hline

        \multirow{16}{=}{User Study}
        & 2-1. & Are there any tools or automation systems you use to make report creation more efficient? If so, what are they? \\
        \cline{2-3}
        & 2-2. & Are there any dissatisfactions or challenges with the tools you use when creating reports? If so, what are they? \\
        \cline{2-3}
        & 2-3. & Is there anything that would be useful to have tool support for in the near future, even if not fully automated? \\
        \cline{2-3}
        & 2-4. & Is there anything you think can be automated in reporting? Conversely, what are the parts that are difficult to automate? \\
        \cline{2-3}
        & 2-5. & Do you have high expectations for using LLMs when creating reports? What do you think about using LLMs for reporting? \\
        \cline{2-3}
        & 2-6. & Have you worked to establish common report formats and templates? If not, what prevents unifying the format? \\
        \cline{2-3}
        & 2-7. & Although report items are somewhat fixed, does the detailed content vary significantly depending on the event? Is it difficult to describe details due to reliance on experience? \\
        \cline{2-3}
        & 2-8. & Is there sufficient feedback on report content from customers or veteran analysts? If so, what kinds of comments have you received? \\
        \cline{2-3}
        & 2-9. & When reviewing reports created by others, what criteria do you use to judge their quality? \\
        \cline{2-3}
        & 2-10. & What do you think are the main challenges in evaluating analysis reports? \\
        \cline{2-3}
        & 2-11. & What do you think is the most important part of an analysis report? \\
        \cline{2-3}
        & 2-12. & What kinds of report items make decision-making easier? \\
        \cline{2-3}
        & 2-13. & Do you feel that the reports you create are sufficient compared to what they ideally should be? If not, what are the main reasons (e.g., time constraints, lack of information)? \\
        \cline{2-3}
        & 2-14. & Is contextual information from the field important in report content? If so, why, and what challenges exist in handling such information? \\
        \cline{2-3}
        & 2-15. & What kinds of information make report creation difficult, either due to their content or the process of collecting and organizing them? \\
        \cline{2-3}
        & 2-16. & What aspects of reporting do you find particularly challenging? \\
        \hline
    \end{tabularx}
\end{table*}

We employ template analysis, which begins with a prior set of themes with interests and iteratively refines them by extending templates by incorporating new themes. 
It is considered that template analysis is useful in domains such as SOC research when prior work has a partial understanding of the concepts that need to be identified in advance~\cite{bushra202299False}. 
First, we conducted an initial round of the coding process based on the descriptions obtained from the literature review and developed an initial set of thematic codes. 
We then analyzed the interview data and, whenever we found new codes that did not match any existing codes, we refined the existing themes by incorporating the new codes. 
The above coding process was performed iteratively three times, with code reviews by SOC practitioners, and codes were finalized through these iterations.
We also note that no new code was obtained after the 14th interview, and hence we judged the saturation of themes. 
Therefore, we did not conduct additional interviews beyond the 15 participants.
In the following, we describe the evaluation criteria for assessing report quality that were derived from the interview results.

\subsubsection{Implicit Evaluation Criteria for Security Report Quality}
\label{sec:implicit-criteria}

\begin{table}[t]
  \centering
  \caption{Evaluation criteria for analysis reports}
  \label{tab:evaluation_perspectives}
  \small
  \setlength{\tabcolsep}{6pt}
  \begin{tabularx}{\textwidth}{X}
    \toprule

    \multicolumn{1}{l}{\textbf{Decision support and action planning for stakeholders}} \\
    \midrule
    (1) Clarity and persuasiveness of the bottom-line conclusion for rapid situation understanding \\
    (2) Quality of concreteness and prioritization in the proposed next actions \\
    (3) Adequacy and clarity of the described customer or business impact \\
    (4) Appropriateness of tailoring the content, tone, and level of detail to the reader’s role and level of understanding \\

    \midrule
    \multicolumn{1}{l}{\textbf{Technical understanding and root-cause clarification of the event for analysts}} \\
    \midrule
    (5) Accuracy and explanatory strength of factual descriptions, including explicit assumptions and constraints \\
    (6) Appropriateness and depth of on-site contextual information usage, such as asset roles and operational background \\
    (7) Adequacy of technical depth in describing the event beyond simple enumeration of logs \\
    (8) Robustness of anomaly identification through the use of multiple analytical approaches, such as trends and comparisons \\

    \midrule
    \multicolumn{1}{l}{\textbf{Accountability and quality assurance of the analysis process for each organization}} \\

    \midrule
    (9) Effectiveness of structured presentation of key analytical elements in supporting accurate understanding of the analysis \\
    (10) Clarity and verifiability of the analysis process through explicit and coherent links between evidence and conclusions \\
    (11) Clarity of analysis scope and responsibility boundaries to facilitate seamless escalation and role demarcation\\

    \bottomrule
  \end{tabularx}
\end{table}

We identify the implicit evaluation criteria that SOC practitioners use in evaluating analysis reports and refine them into explicit evaluation criteria. 
Although most of the organizations we interviewed did not maintain formal documents of evaluation criteria for analysis reports, we found that several veteran SOC practitioners, such as Tier-2–level analysts and recipient-side managers, have similar criteria when they evaluate analysis reports.
Moreover, most comments pointed out the quality of the report content instead of its formal quality. 
We present the resulting evaluation criteria and their related comments in accordance with the three categories, including the items that should be included in the report.

\paragraph{}\textbf{Evaluation Criteria for Decision Support and Action Planning}

\textbf{Criterion (1): Clarity and persuasiveness of the bottom-line conclusion for rapid situation understanding.}
The participants emphasized that analysis reports should present a clear and persuasive conclusion that enables readers to take and decide next actions. 
In particular, analysis reports are expected to explicitly document whether security alerts are valid and urgent, the recommended next actions against the alerts, and the reasons that support these judgments. 
However, the participants noted that many analysis reports merely enumerate factual descriptions without documenting the above information.
\emph{``High-quality reports clearly state whether an alert should be treated as a true positive or not, how urgent it is, and why that judgment was made. Without that bottom line, it is hard to decide the next step quickly.''} –- UserID~1.
\emph{``For decision-making, the key is whether the report clearly explains what the problem is, why it matters, and what conclusion we should draw from the analysis.''} –- UserID~3.

\textbf{Criterion (2): Quality of concreteness and prioritization in the proposed next actions.}
The participants noted that analysis reports should not end with analysis results but should clearly outline the next actions with their priorities. 
Effective analysis reports enable readers to distinguish between actions that must be taken immediately and those that are optional or conditional, thereby allocating resources appropriately. 
In contrast, analysis reports that simply list insights without actionable guidance are insufficient for operational use.
\emph{``It is important to clearly state what should be done next and to indicate priorities, for example, whether an action is mandatory or should be handled if resources allow.''} –- UserID~9.
\emph{``Readers need to understand not only what issues exist, but also what actions they themselves are expected to take in response to those issues.''} –- UserID~6.

\textbf{Criterion (3): Adequacy and clarity of the described customer or business impact.}
The participants regarded it as essential that analysis reports clearly describe how each security alert affects the customer’s business and operations from the customer’s perspective. 
Analysis reports that focus solely on technical details without considering business impact are often difficult for readers to judge whether the underlying security alerts are acceptable. 
Quantifying the impact is also valuable for decision-making.
\emph{``If the impact on business or operations is clearly described or even roughly quantified, it becomes much easier to explain the situation and move forward.''} –- UserID~11.
\emph{``Ultimately, customers want to know what caused the issue and whether the situation is truly acceptable. If that is unclear, the report leaves them uneasy.''} –- UserID~14.

\textbf{Criterion (4): Appropriateness of tailoring the content, tone, and level of detail to the recipient’s role and level of understanding.}
The participants highlighted that it is important to adjust the content, including its technical details, to readers' roles and expertise in order to reduce cognitive cost. 
It also enables the readers to understand whether security alerts are urgent. 
The participants noted that reports written in a uniform style often fail to bridge the gap between frontline analysts and stakeholders who make decisions, leading to misunderstandings and delayed responses.
\emph{``Readability is critical. Each recipient has a different level of understanding, so the explanation and terminology need to be adjusted accordingly.''} –- UserID~2.
\emph{``What matters most is narrowing the gap in sense of urgency between the frontline and decision-makers. Reports should reflect the context and discussions that led up to the analysis.''} –- UserID~10.

\paragraph{}\textbf{Evaluation perspectives for technical understanding and root-cause clarification of the event}

\textbf{Evaluation criterion (5): Accuracy and explanatory strength of factual descriptions, including explicit assumptions and constraints.}
The participants emphasized that analysis reports must precisely describe what occurred during the event and provide explanations to convincingly justify the conclusion. 
The participants highlighted that explicitly stating the assumptions, analytical grounds, and constraints underlying the analysis is important beyond factual descriptions. 
Analysis reports that lack these contents are often difficult for readers to trust the conclusion. 
\emph{``The most important thing is to accurately describe what actually happened and to explain it in a way that the reader can reasonably accept.''} –- UserID~7.
\emph{``If the assumptions, grounds, or constraints behind the analysis are unclear, it becomes hard to judge whether the situation has really been understood correctly.''} –- UserID~4.

%
\textbf{Evaluation criterion (6): Appropriateness and depth of on-site contextual information usage, such as asset roles and operational background.}
The participants noted that proper interpretation of security alerts needs operational context, including asset roles and their position within operational processes. 
Unless such context is present, analysts are often unable to determine whether each security alert is a genuine issue or a false positive. 
The participants also noted that insufficient operational context frequently leads to overly ambiguous conclusions.
\emph{``Operational context is indispensable. Depending on the role of the asset and how it is used, the priority and meaning of the alert can change completely.''} –- UserID~13.
\emph{``If information such as asset roles, whether the traffic is business-related, or results of on-site interviews is missing, we often cannot clearly determine whether an alert is a false positive.''} –- UserID~5.

%

\textbf{Evaluation criterion (7): Adequacy of technical depth in describing the event beyond simple enumeration of logs.}
Rather than merely listing communication records and log entries, analysis reports should provide technically meaningful explanations of the event, such as what they are used for. 
Several participants expressed frustration that interpreting raw technical data is often insufficient.
\emph{``SOC reports often just list IP addresses that communicated, and we have to re-investigate where those systems are and what they are used for.''} –- UserID~6.
\emph{``If the report also explained the meaning of the communication, it would significantly reduce the investigation workload on our side.''} –- UserID~11.

%
\textbf{Evaluation criterion (8): Robustness of anomaly identification through multiple analytical approaches, such as trends and comparisons.}
The participants underscored that identifying anomalies using multiple analytical perspectives is important, including temporal trends, comparisons with historical data, and correlations across different logs. 
It is considered that analysis reports that rely solely on isolated log entries are insufficient for understanding whether observed behavior is truly an anomaly. 
The participants noted that temporal and comparative analyses are still underutilized in many current analysis reports.
\emph{``What we really want to know is what has changed compared to the past, but that perspective is often missing in reports.''} –- UserID~9.
\emph{``Showing insights derived from differences in trends or combinations of logs would make anomalies much clearer.''} –- UserID~3.

\paragraph{}\textbf{Evaluation perspectives for accountability and quality assurance of the investigation and analysis process}

\textbf{Evaluation criterion (9): Effectiveness of structured presentation of key analytical elements in supporting accurate understanding of the analysis.} 
The participants emphasized that structuring and explicitly presenting key analytical elements, such as hypotheses, examined evidence, and interim judgments, is essential for ensuring that readers can precisely understand the analysis. 
Although standardized formats and procedures exist, the participants noted that differences in security alerts and analysts' experiences often result in variations in how these elements are clear, thereby sometimes leading to inconsistent qualities across analysis reports.
\emph{``There is noticeable variation in report quality, and it often becomes dependent on the experience of individual analysts, especially in how clearly the analysis is structured.''} –- UserID~4.
\emph{``Even if formats are standardized, the way key analytical points are organized and explained still differs by analyst and by case.''} –- UserID~1.

\textbf{Evaluation criterion (10): Clarity and verifiability of the analysis process through explicit and coherent links between evidence and conclusions.}
It is considered critical that analysis reports clearly document the reasoning process by explicitly linking evidence to conclusion, so that a third party can trace and verify how judgments are reached. 
The participants pointed out that analysis reports often fail to sufficiently document how alternative explanations, such as false positives, were examined, thereby undermining confidence in conclusion and the quality of the analysis. 
\emph{``It is important that there are no logical gaps or contradictions in how the analysis results lead to the conclusions, and that a third party could reproduce the same reasoning.''} –- UserID~7.
\emph{``Especially for determining whether an alert is a false positive, the evidence and reasoning must be clearly connected. If that part is weak, the quality of the entire report appears low.''} –- UserID~11.

\textbf{Evaluation criterion (11): Clarity of analysis scope and responsibility boundaries to facilitate seamless escalation and role demarcation.} 
The participants emphasized that effective analysis reports clearly define the scope of investigation performed by SOC practitioners and explicitly state responsibility boundaries for subsequent actions. 
High-quality reports do not attempt to exhaustively cover all aspects of an incident: instead, they clarify what has been analyzed, what conclusions can be reasonably drawn, and where responsibility is intentionally handed over to other stakeholders, such as CSIRT teams or customer-side organizations. 
This clarity facilitates seamless escalation, avoids redundant analysis, and enables efficient collaboration across roles and organizational layers.
\emph{``It is essential to clearly state what has been analyzed and what our assessment is, and then promptly escalate the case once it goes beyond our scope. Defining the boundary and handing it over early is crucial for enabling rapid and effective response.''} –- UserID~12.
\emph{``Because roles and scopes are clearly defined—both between SOC and CSIRT and across Tier 1, Tier 2, and senior analysts—we can escalate without hesitation and avoid over-analyzing matters that fall outside our responsibility.''} –- UserID~11.

\begin{table}[t]
  \centering
  \caption{Representative examples of categories, report elements, and checklist items with mapped evaluation criteria}
  \label{tab:checklistsample}
  \small
  \setlength{\tabcolsep}{6pt}
  \begin{tabularx}{\textwidth}{p{0.32\textwidth} X}
    \toprule
    Category / report element 
      & Example checklist item with mapped evaluation criteria \\
    \midrule

    \multicolumn{2}{l}{\textbf{Decision support and action planning}} \\
    \midrule

    Impact Assessment
      & Are the on-site impacts, severity level, and risk evaluation described with justification? --- (3)  \\

    Confirmation Requests
      & Are specific items or questions to be confirmed by the recipient regarding this event described? --- (2),(4)\\

    Response Actions
      & Are specific actions taken, their necessity, and the effects and evaluation methods described? ---  (1),(2)\\

    \midrule
    \multicolumn{2}{l}{\textbf{Technical understanding and root cause clarification of the event}} \\
    \midrule
    Event Description \& Interpretation
      & Does the report draw a concrete conclusion based on the analysis of the observed event? --- (5),(7) \\

    Root Cause
      & If the incident is assumed to be an operational effect, is this explained? --- (6),(7),(8)  \\

    Incident Source / Impacted Systems
      & Is the origin device clearly identified along with its role, importance, and suspiciousness? --- (6),(7)  \\

    \midrule
    \multicolumn{2}{l}{\textbf{Accountability and quality assurance of the investigation and analysis process}} \\
    \midrule
    Investigation \& Analysis Method
      & Are the analysis methods, decision criteria, and viewpoints explained in a multi-faceted manner? --- (9),(11) \\

    Evidence \& Supporting Data
      & Is evidence provided to support each observation or response, with source and acquisition method? --- (10) \\

    Analysis Status
      & Is the current analysis progress and the next steps clearly stated? ---(11)  \\
    
    \bottomrule
  \end{tabularx}
\end{table}

\subsection{Development of the Analyst-wise Checklist Draft}
The Analyst-wise Checklist was developed by systematically integrating two sources, i.e., 1) the structural elements that should be documented in analysis reports, which were extracted through a literature review, and 2) the implicit evaluation criteria used in practice by SOC analysts, which were obtained through interviews. 
The integrated item set was then finalized through expert review and application tests. This approach is also widely known as a general method for constructing evaluation checklists~\cite{hales2008development}. 

\paragraph{Integration Method}

To construct the checklist items, we first map, for each structural element of analysis reports, the evaluation criteria derived from SOC practitioners’ challenges described in Section~\ref{sec:implicit-criteria}. 
We then define concrete evaluation items by drawing on interview comments and descriptions of best practices extracted from the literature review.

For designing these items, we considered whether actions suggested by the checklist are feasible: including all the analysts' concerns in the checklist may exceed a typical authority and resources of SOCs~\cite{rose2022co}. 
The checklist items were therefore refined to remain actionable under realistic environments.
For example, for the evaluation criterion (6), the participants repeatedly pointed out that the lack of local context often prevents them from deciding whether a security alert is a false positive. 
However, it is often unrealistic for SOCs to always obtain complete and latest knowledge for every environment~\cite{socchallenge}. 
The checklist requires obtaining sufficient context of SOC practitioners, and asks whether they gather and leverage knowledge for decision-making. 
The above design reflects the practical roles for SOCs, including managers, and therefore, will make report interpretation certain.
To ensure applications of the checklist into various domains, each checklist item is designed with generalized granularity, thereby avoiding overly domain-specific criteria. 
The checklist is expected for SOC practitioners to select only items relevant to each analysis report rather than the use of the entire checklist.

These integration steps were primarily conducted by two researchers, including the first author. 
We also used a general-purpose LLM (gpt-4.1): specifically, given the organized evaluation perspectives and structural elements, we prompted the model to “enumerate candidate checklist items for each perspective,” and used the generated suggestions to expand and refine candidate items and their wording.
Before the expert review, the draft items were discussed among the authors and refined to retain only promising candidates.

\subsection{Refinement through Expert Review and Evaluation Tests} 

The checklist is finalized by expert review by SOC practitioners and application tests. 
This approach is a general method for constructing evaluation checklists~\cite{hales2008development}. 
First, three SOC practitioners reviewed the checklist and provided feedback on whether it is practical, whether it helps prevent omissions during review, and whether its contents are consistent with operations that should be performed within a SOC. 
Specifically, we received comments grounded in actual SOC practice: “It is not the SOC’s role to prescribe concrete response actions, but they are expected to provide assessment; therefore, such strictness should not be directly reflected in the checklist items.” 
Based on these results, we refined the wording of several items and conducted a second round of review with reference to the Delphi method~\cite{rose2022co}. 
We then selected as consensus items those that at least two of the three SOC practitioners evaluated as useful.

As a final validation step, one of the three SOC practitioners and one of the authors, who have more than seven years of SOC-related experience, independently applied the checklist to ten real-world analysis reports for security alerts. 
For each checklist item, they rated it on a five-point Likert scale. 
Then, for items where their ratings substantially differ from each other, e.g., one judged the item to be largely satisfied while the other judged it to be not satisfied, they reconciled their interpretations and refined the item descriptions and wording accordingly.

We confirmed that for more than 60\% of the checklist items, the two raters agreed on the direction of judgment, i.e., whether the item was satisfied or not, and we then finalized the checklist. 
Table~\ref{tab:checklistsample} shows a subset of the checklist with mappings to the evaluation criteria. 
The complete checklist and checklist item types are provided in Table~\ref{tab:checklistsample_type} and Table~\ref{tab:checklistitems}.

\begin{table*}[tbp]
    \caption{Analyst-wise Checklist Item Type Sample. This checklist summarizes essential items that should be confirmed in a typical security incident analysis report. An additional column labeled ``Optional'' indicates whether the item is conditionally required depending on the incident context—for example, some items may be primarily addressed by a dedicated Security Incident Response Team (SIRT).}
    \label{tab:checklistsample_type}
    \footnotesize
    \centering
    \resizebox{\textwidth}{!}{
    \begin{tabular}{|
        p{4.2cm}
        |p{3.2cm}
        |p{7.2cm}
        |c|}
        \hline
        \textbf{Report Content Category} &
        \textbf{Checklist Item Type} &
        \textbf{Details} &
        \textbf{Optional} \\
        \hline

        \multirow{7}{=}{Technical understanding and root cause clarification of the event}
        & Basic Alert Information
        & Basic alert metadata such as timestamp, alert name, alert ID, and the detecting system that triggered the event.
        & No \\
        \cline{2-4}

        & Event Description \& Interpretation
        & A clear description and interpretation of the observed event, including what happened, how it was detected, and why it is considered relevant.
        & No \\
        \cline{2-4}

        & Root Cause
        & Identification and explanation of the underlying technical or human causes that led to the incident.
        & No \\
        \cline{2-4}

        & Incident Source
        & The origin of the incident, such as the attack source, initiating asset, user, or external entity involved.
        & No \\
        \cline{2-4}

        & Impacted Systems
        & The systems, assets, services, or environments affected by the incident or potentially impacted.
        & No \\
        \cline{2-4}

        & Communication Details
        & Details of suspicious or relevant communications, including endpoints, protocols, destinations, content characteristics, and frequency.
        & No \\
        \cline{2-4}

        & Vulnerability Information
        & Information on known or suspected vulnerabilities related to the incident, if applicable.
        & No \\
        \hline

        \multirow{5}{=}{Decision support and action planning}

        & Impact Assessment
        & An assessment of the operational impact, severity, business relevance, and associated risk of the incident.
        & No \\
        \cline{2-4}

        & Confirmation Requests
        & Specific items, assumptions, or questions that require confirmation from on-site staff, system owners, or other relevant teams.
        & No \\
        \cline{2-4}

        & Response Actions
        & Actions taken or recommended in response to the incident, including containment, mitigation, or monitoring measures.
        & Yes \\
        \cline{2-4}

        & Recurrence Prevention \& Lessons Learned
        & Post-incident considerations such as preventive measures, control improvements, and lessons learned to avoid recurrence.
        & Yes \\
        \hline

        \multirow{3}{=}{Accountability and quality assurance of the investigation and analysis process}
        & Investigation \& Analysis Methods
        & The methods, tools, data sources, and analytical procedures used during the investigation and analysis of the event.
        & No \\
        \cline{2-4}

        & Analysis Status
        & The current analysis status of the incident, including whether it is ongoing, unresolved, or concluded, and the rationale for that status.
        & No \\
        \cline{2-4}

        & Evidence \& Supporting Data
        & Preservation and documentation of supporting evidence, such as logs, alerts, configuration data, or artifacts used in the analysis.
        & Yes \\
        \cline{2-4}

        & Related Policies, Guidelines \& Standards
        & Relevant internal policies, operational guidelines, or external standards applicable to the incident or its handling.
        & Yes \\

        \hline

    \end{tabular}
    }
\end{table*}

\begin{table*}[tpb]
  \caption{Analyst-wise Checklist Items. Each checklist item specifies a concrete review question under a checklist item type.}
  \label{tab:checklistitems}
  \footnotesize
  \centering
  \resizebox{\textwidth}{!}{
  \begin{tabular}{|p{3.5cm}|p{14cm}|}
    \hline
    \textbf{Checklist Item Type} & \textbf{Checklist Item} \\
    \hline

    \multirow{1}{=}{Basic Alert Information}
    & Does the report include basic information about the alert under analysis, such as its content, time of occurrence, and location? \\
    \hline

    \multirow{2}{=}{Analysis Status}
    & Is the current analysis progress and the next steps clearly stated? \\
    \cline{2-2}
    & Has it been confirmed whether the observed event is still ongoing and whether any related events are occurring? \\
    \hline

    \multirow{4}{=}{Event Description \& Interpretation}
    & Does the report draw a concrete conclusion based on the analysis of the observed event? \\
    \cline{2-2}
    & Are the key messages of the report and the rationale for its conclusions sufficiently explained? \\
    \cline{2-2}
    & If the supporting information is insufficient, does the report indicate that appropriate inquiries have been made to the recipient side? \\
    \cline{2-2}
    & Is the event interpreted and explained in line with the on-site operational context? \\
    \hline

    \multirow{11}{=}{Investigation \& Analysis Methods}
    & Does the report describe the methods used to assess the likelihood of an attack and its potential impact for the observed event? \\
    \cline{2-2}
    & Are the analysis methods, decision criteria, and viewpoints explained in a multi-faceted manner?\\
    \cline{2-2}
    & Has the event been checked by comparing its communications and device behavior with usual traffic and behavior patterns? \\
    \cline{2-2}
    & Does the report verify whether the observed behavior actually matches the attack pattern that the alert is designed to detect? \\
    \cline{2-2}
    & Have past alerts been investigated as well? \\
    \cline{2-2}
    & In addition to the alert in question, have related alerts in the surrounding time period been examined? \\
    \cline{2-2}
    & Has it been checked whether there are any anomalies in the temporal sequence of communications and behaviors before and after the alert-triggering event? \\
    \cline{2-2}
    & Based on the communications and behaviors associated with the alert, does the report conduct a detailed investigation and risk assessment to determine whether any abnormal actions succeeded and to assess the likelihood of an attack? \\
    \cline{2-2}
    & Has it been verified whether the observed actions or communications deviate from the expected role and normal operation of the device or user? \\
    \cline{2-2}
    & Have external reputation and vulnerability information sources been consulted, and has their consistency with the observed event been evaluated? \\
    \cline{2-2}
    & Has the event been evaluated in light of site-specific allowed operations and prohibited (“NG”) conditions? \\
    \hline

    \multirow{5}{=}{Root Cause}
    & Has the root cause of the alert been identified and documented? \\
    \cline{2-2}
    & Does the report explain the analysis that led to the identification of the cause? \\
    \cline{2-2}
    & Is it distinguished whether the cause is human (manual operation, maintenance) or system-related (automatic/periodic processing, failure, malfunction, etc.)? \\
    \cline{2-2}
    & If the cause may be an attack, is the type of attack described concretely? \\
    \cline{2-2}
    & If the incident is assumed to be an operational effect, is this explained? \\
    \hline
    
    \multirow{7}{=}{Incident Source}
    & Is detailed information about the source of the alert (e.g., the device’s on-site name and role) documented? \\
    \cline{2-2}
    & Has the importance of the source (device/user) been assessed? \\
    \cline{2-2}
    & Has the degree of suspiciousness of the source been evaluated? \\
    \cline{2-2}
    & Does the report describe any changes in the state of the source? \\
    \cline{2-2}
    & Does the report describe the current status of the source? \\
    \cline{2-2}
    & Has it been checked whether the communications performed by the device or user are appropriate given its intended role? \\
    \cline{2-2}
    & If the device or user is unknown, does the report provide a reasonable hypothesis about its possible identity or role? \\
    \hline

    \multirow{5}{=}{Impacted Systems}
    & Is information about the alert’s communication destination documented in sufficient detail and without omissions? \\
    \cline{2-2}
    & Has the importance (criticality) of the destination resource been assessed? \\
    \cline{2-2}
    & Does the report describe any changes in the state of the communication destination? \\
    \cline{2-2}
    & Does the report describe the current status of the communication destination? \\
    \cline{2-2}
    & If the destination is an unknown device, does the report provide a reasonable hypothesis about its identity or role? \\
    \hline

    \multirow{4}{=}{Communication Details}
    & Is the content of the alert-related communication documented in detail, including responses and whether each communication succeeded or failed? \\
    \cline{2-2}
    & Does the report explain how the communication relates to on-site operations? \\
    \cline{2-2}
    & Is the abnormality of the communication and the likelihood of an attack described and evaluated concretely? \\
    \cline{2-2}
    & Does the report include information about the network segments (e.g., VLAN IDs, segment names, and their respective roles)? \\
    \hline

    \multirow{3}{=}{Impact Assessment}
    &  Are the on-site impacts, severity level, and risk evaluation described with justification? \\
    \cline{2-2}
    & Does the report concretely describe the potential impact if the event were an actual attack? \\
    \cline{2-2}
    & Are the criteria and rationale used for the risk assessment explicitly stated? \\
    \hline

    \multirow{3}{=}{Confirmation Requests}
    & If confirmation by the recipient is required, is the necessity of such confirmation and its rationale sufficiently explained? \\
    \cline{2-2}
    & Are specific items or questions to be confirmed by the recipient regarding this event described? \\
    \cline{2-2}
    & Are the requested confirmations or investigations realistically feasible for the recipient side? \\
    \hline

    \multirow{3}{=}{Response Actions}
    & Is it clearly stated whether response actions are required, and is the rationale for this necessity sufficiently explained? \\
    \cline{2-2}
    & Are specific response actions for the current event described? \\
    \cline{2-2}
    &  Are specific actions taken, their necessity, and the effects
and evaluation methods described? \\
    \hline

    \multirow{4}{=}{Recurrence Prevention \& Lessons Learned}
    & Are concrete recurrence-prevention or improvement measures described? \\
    \cline{2-2}
    & Does the report include the expected effects of these preventive/improvement measures and how they will be evaluated or monitored? \\
    \cline{2-2}
    & Are the proposed recurrence-prevention or improvement measures realistic and actionable? \\
    \cline{2-2}
    & Does the report articulate lessons learned from the event and future response policies or courses of action? \\
    \hline

    \multirow{2}{=}{Evidence \& Supporting Data}
    &  Is evidence provided to support each observation or response, with source and acquisition method? \\
    \cline{2-2}
    & Are the methods and sources for obtaining the evidence described? \\
    \hline

    \multirow{3}{=}{Vulnerability Information}
    & Has known vulnerability information relevant to the event been collected? \\
    \cline{2-2}
    & Is the collected vulnerability information linked to and explained in relation to the analysis findings? \\
    \cline{2-2}
    & Does the report explain whether and how the identified vulnerabilities may apply to the observed event? \\
    \hline

    \multirow{2}{=}{Related Policies, Guidelines \& Standards}
    & Does the report identify any policies, guidelines, or industry standards that may have been violated? \\
    \cline{2-2}
    & Has the risk and impact of such non-compliance been assessed? \\
    \hline

  \end{tabular}
  }
\end{table*}

\subsection{Answer to RQ1} \label{sec:answersrq1}
Based on the findings from the literature review and the semi-structured interviews, we define the evaluation criteria of analysis reports as the following three standpoints:
1) Decision support and action planning for stakeholders, determining what actions are necessary and why; 
2) Technical understanding and root-cause clarification of the event for analysts, providing accurate descriptions of system behavior and its operations;
3) Accountability and quality assurance of the analysis process for each organization, ensuring that conclusions of the report are transparently linked to its evidence and reasoning.
The quality of analysis reports for SOCs is judged by evaluation criteria consisting of multiple perspectives of SOC practitioners. 
Our Analyst-wise Checklist contains these evaluation criteria. 

\endgroup

\begingroup

\section{Proposed Method: MESSALA} \label{sec:proposed} 

In this section, we propose MESSALA, which evaluates analysis reports with feedback, using the Analyst-wise Checklist in the previous section. We first explain design motivation and the key ideas behind MESSALA, and then describe its details.

\subsection{Design Motivation and System Requirements}

Our motivation for MESSALA is to incorporate the cognition and domain knowledge of veteran SOC practitioners into LLM-based evaluation.
The Analyst-wise Checklist is utilized to realize this idea. 
However, using the checklist alone is still insufficient because LLM-based judgments of texts often differ from evaluations of human experts~\cite{murugadoss2025evaluating}. 
Indeed, when we conducted post-interviews with five of the fifteen participants in the interviews with respect to the evaluation of analysis reports by an LLM using the Analyst-wise Checklist, we obtained feedback similar to that reported in prior
work~\cite{murugadoss2025evaluating}
, such as:\emph{``The feedback remains largely superficial and does not appear to provide a sufficiently in-depth evaluation of the report.'', 
``The comments avoid direct wording, which makes it difficult to translate the feedback into concrete actions.''}

Therefore, to provide appropriate evaluation results, an LLM must judge a report consistently with evaluations of veteran SOC practitioners and synthesize these judgments into feedback~\cite{zhou2024llm}. 
To do this, we formulate two elements, i.e., numerical scores for quantitative evaluation and feedback for qualitative evaluation, for the evaluation of analysis reports, which correspond to RQ2 and RQ3. 
MESSALA is designed to address these questions. 

\subsection{Key Idea}
The key idea of MESSALA is to imitate the cognitive process used by SOC practitioners~\cite{zhong2016automate} when evaluating analysis reports, as this enables LLMs to generate outputs similar to humans~\cite{saha2025learning,chu2025think}. 
To realize this idea, we introduce two new methods, \textit{Granularization Guideline} and \textit{Multi-perspective Evaluation LLM}, in addition to the Analyst-wise Checklist. 
MESSALA can then generate evaluation results and feedback identical to those of veteran SOC practitioners for analysis reports. 
We describe each method below. 

The Granularization Guideline steers the LLM’s inference with expertise and practical feedback of veteran SOC practitioners by translating the Analyst-wise Checklist into actionable evaluations. 
This enables an LLM to focus on the specific parts that require detailed evaluations of veteran SOC practitioners, in analysis reports.


On the other hand, the Multi-perspective Evaluation LLM is a new architecture to obtain human SOC practitioners’ judgments and their effective feedback with respect to evaluations of analysis reports.
This architecture integrates two evaluation processes, i.e., \textit{High-level Evaluation LLM} as an initial evaluation based on the superficial information of the given report with a simple prompt and \textit{In-depth Evaluation LLM} as a detailed evaluation by taking its context using the Granularization Guideline into account. 
A multi-perspective evaluation is instantiated by integrating both results. 
This design imitates the cognitive process of human text recognition~\cite{wharton1991overview}: humans first activate knowledge fragments through superficial features, and then integrate their contexts to understand the meaning of the given text.
By reproducing this human cognitive process within the architecture, the Multi-perspective Evaluation LLM enables LLMs to more deeply understand the Analyst-wise Checklist as human SOC practitioners' knowledge.

\subsection{Detail of the Method} \label{sec:detailsofMethod}

\begin{figure}[t]	
	\centering
	\includegraphics[width=\linewidth]
    {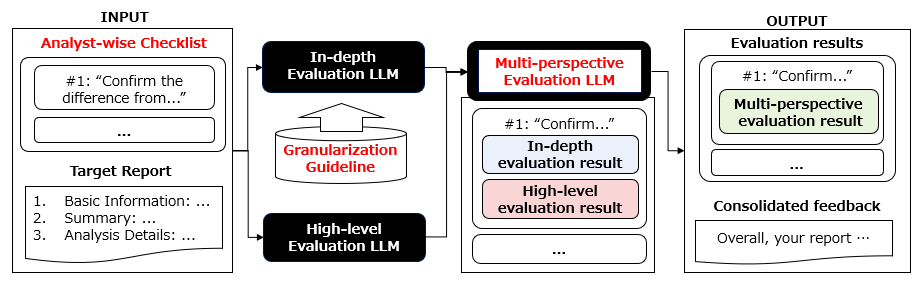}
	\caption{Overview of MESSALA. 
    The top module runs two parallel processes:
High-level Evaluation LLM using only the checklist and an In-depth Evaluation LLM using the Granularization Guideline.
Their outputs are integrated by Multi-perspective Evaluation LLM to return an output.
    }
	\label{fig:proposedmethod}
\end{figure}

The overall framework of MESSALA is illustrated in Fig.~\ref{fig:proposedmethod}.
MESSALA consists of three components: Analyst-wise Checklist, Granularization Guideline, and multi-perspective evaluation. 
Since the Analyst-wise Checklist was explained in the previous section, the following will describe the remaining two constituent components.

\subsubsection{Granularization Guideline}


The Granularization Guideline specifies, for each category
of the Analyst-wise Checklist, contextual information that should be documented. This
guideline represents an abstraction rule to guide judgments of LLMs, whereas the
Analyst-wise Checklist represents specific items for evaluations of analysis reports.
Representative examples of categories in The Granularization Guideline are shown in
Table~\ref{tab:guideline}. For example, the category of “Hypothesis Validation" focuses on guiding LLMs
to identify evidence supporting the conclusion of the given reports. These categories
are designed to evaluate whether sufficient context for the analysis is provided, as well
as whether the analysis methods described in the reports align with those of SOC
practitioners. They also evaluate whether hypotheses are reasonably justified.

If the checklist item is “If the incident is assumed to be an operational effect, is this explained?”, the In-depth Evaluation LLM refers to the Hypothesis Validation category in the guideline. It first extracts the operational hypothesis described in the report and then granularizes the item into report-specific confirmation points. In this case, the confirmation points include whether the maintenance activity is concretely described, whether the timing of the anomalous communications is consistent with the maintenance notice, whether observations from other known devices support the maintenance hypothesis, and whether protocol-level details are explained as plausible during maintenance. 
The role of the guideline in In-depth Evaluation LLM is illustrated in Fig.~\ref{fig:proposedmethod_guide}, and Fig.~\ref{fig:gra_sample} shows a sample sanitized report and the resulting detailed confirmation points generated for this checklist item.

The Granularization Guideline is designed to granularize Analyst-wise Checklist items for a given analysis report into concrete confirmation points. 
Its design follows prior works on SOC practices~\cite{kersten2023give,kerstenfield2025tier1analyst,turcotte2025automated,zhong2016automate}.
Based on this design, categories in the guideline were iteratively refined through evaluations of real-world analysis reports using the Analyst-wise Checklist items in Section~\ref{sec:EvaluationwithAnalyst-Wise}, and then were finalized by confirmation with two SOC practitioners from the authors’ affiliations.
By leveraging the Granularization Guideline, the In-depth Evaluation LLM can evaluate not only the lack of content in analysis reports, but also whether the contexts of the reports are consistent. 
Finally, the In-depth Evaluation LLM forwards the results to the Multi-perspective Evaluation LLM described later.
Prompt examples for each category of Granularization Guideline are shown in Fig.~\ref{fig:guide_prompt_all}.

\begin{table}[t]
  \caption{Overview of categories defined in the Granularization Guideline.}
  \label{tab:guideline}
  \centering
  \begin{tabularx}{\linewidth}{p{3cm} X}
    \hline
     \textbf{Guideline Category} & \textbf{Rationale with Mapped Evaluation Perspectives} \\
    \hline
    Hypothesis Validation &
      Ensures that analytical conclusions are logically traceable and justified by evidence, preventing unsupported or arbitrary judgments. -- (5),(10) \\
    \hline
    Basic Alert Information &
      Ensures that the fundamental facts of the alert are clearly stated so that the analysis is interpretable and actionable. -- (1),(9) \\
    \hline
    Pattern and Comparison Analysis &
      Enables analysts to assess the incident’s significance by contextualizing it against known benign or malicious patterns. -- (8),(10) \\
    \hline
    Detailed Communication Analysis &
      Allows verification of whether observed communications meaningfully indicate abnormal or malicious behavior. -- (5),(7),(10) \\
    \hline
    Temporal and Situational Context Analysis &
      Prevents misinterpretation by considering event sequences and surrounding activities rather than isolated observations. -- (8),(10),(11) \\
    \hline
    Operational Context Utilization &
      Grounds the analysis in real-world system usage and constraints, reducing false conclusions detached from on-site operations. -- (4),(6) \\
    \hline
    Risk, Impact, and Severity Assessment &
      Supports informed decision-making by clarifying the practical consequences and urgency of the incident. -- (1),(2),(3) \\
    \hline
    Threat Scenario Specificity &
      Helps stakeholders understand plausible attack progressions by articulating concrete and coherent threat scenarios. -- (7),(8),(10) \\
    \hline
  \end{tabularx}
\end{table}

\begin{figure}[t]	
	\centering
	\includegraphics[scale=0.7]
    {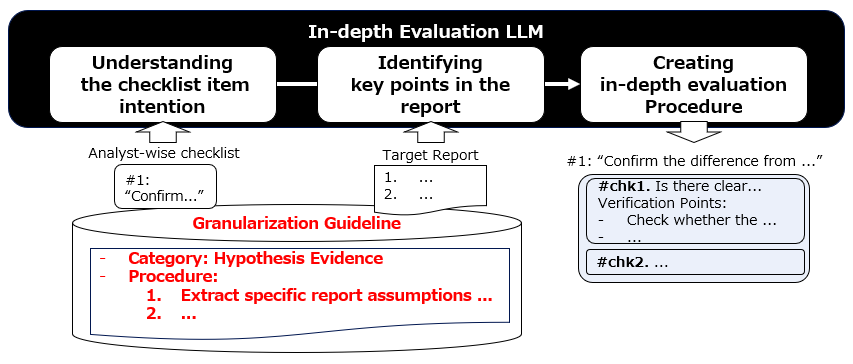}
	\caption{Illustration of the Granularization Guideline in use.
The guideline illustrates how checklist items are broken down into concrete evaluation steps for each category.
 }
	\label{fig:proposedmethod_guide}
\end{figure}

\begin{figure}[t]	
	\centering
	\includegraphics[scale=0.7]
    {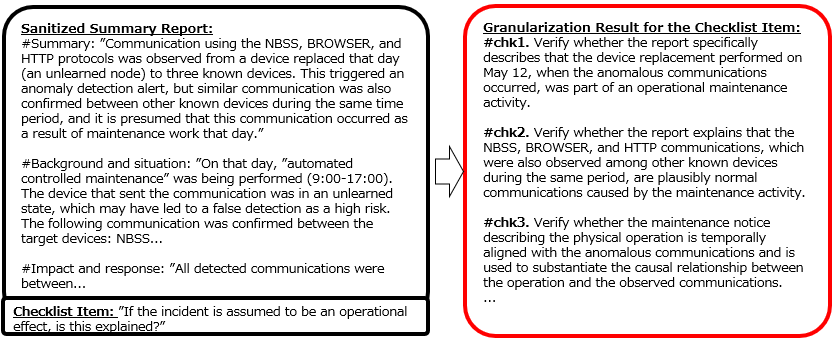}
	\caption{A sample sanitized report and checklist item, and the resulting report-specific confirmation points generated through granularization.
 }
	\label{fig:gra_sample}
\end{figure}

\begin{figure}[tbp]
  \centering
  \begin{subfigure}{\linewidth}
    \centering
\includegraphics[width=\linewidth]{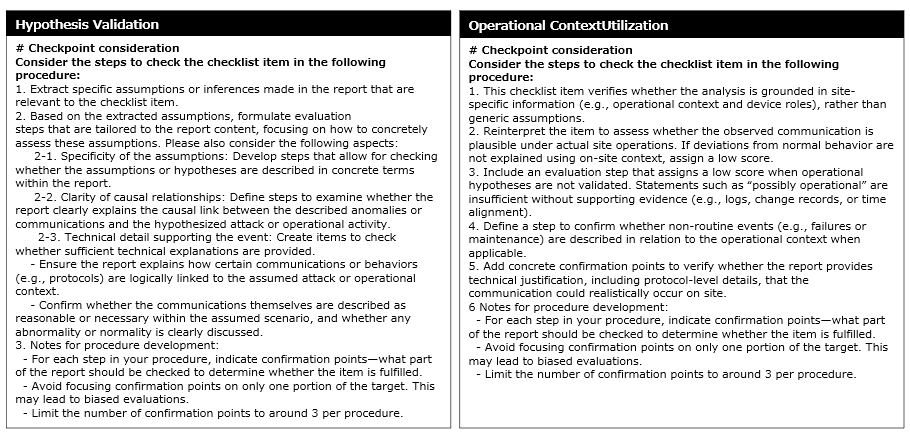}  \end{subfigure}

  \vspace{2mm}

  \begin{subfigure}{\linewidth}
    \centering
\includegraphics[width=\linewidth]{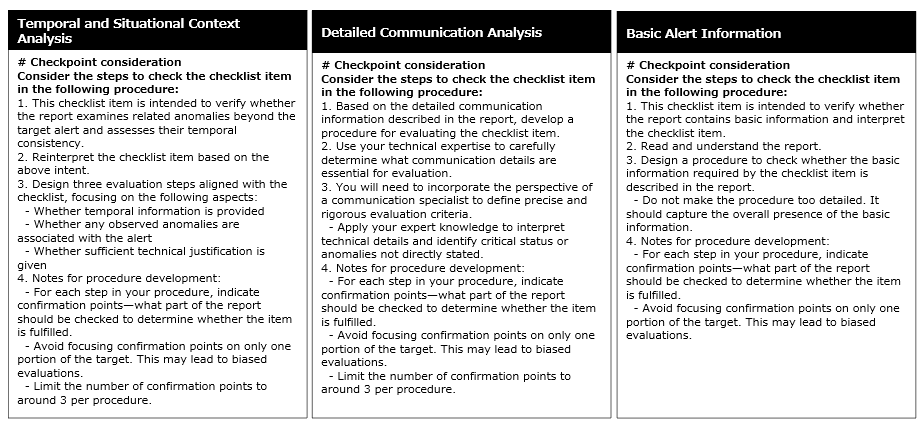}  \end{subfigure}

  \vspace{2mm}

  \begin{subfigure}{\linewidth}
    \centering
\includegraphics[width=\linewidth]{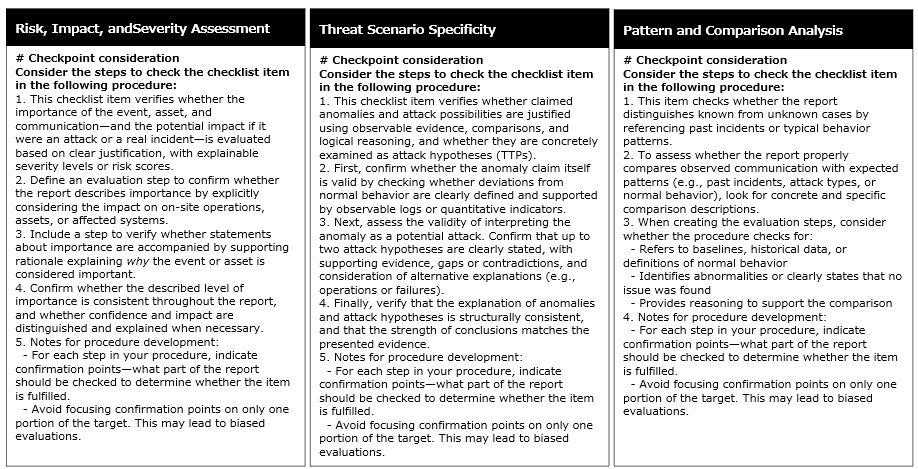}
  \end{subfigure}

	\caption{Overview of Prompts by Category in the Granularization Guideline.}
	\label{fig:guide_prompt_all}
\end{figure}

\subsubsection{Multi-perspective Evaluation LLM with High-level and In-depth Evaluations} \label{sec:multi-perspectiveevaluation}

\begin{figure*}[t]	
	\centering
	\includegraphics[width=\linewidth]{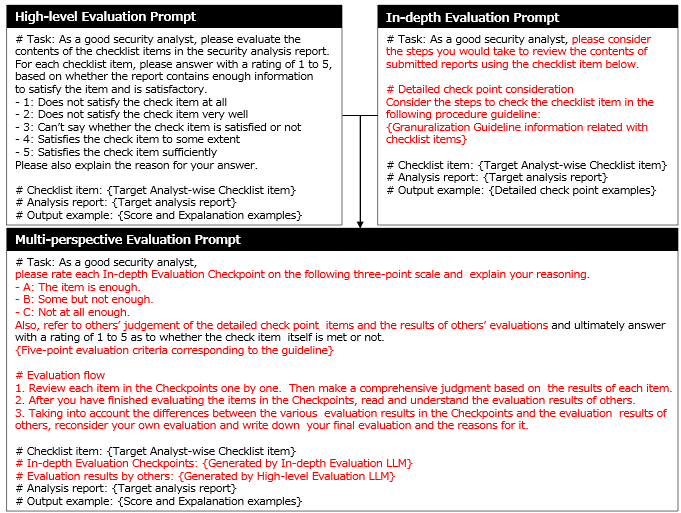}
	\caption{Overview of Prompts for Granularization and Evaluation. The red text indicates the distinctive characteristics of each prompt.
    }
	\label{fig:prompt_all}
\end{figure*}

Our Multi-perspective Evaluation LLM integrates the high-level and in-depth evaluations described above. 
We first describe each evaluation, and then describe their integration. 

\textbf{High-level Evaluation LLM:}
 \label{sec:highlevelevaluation}
This allows an LLM to rely on simple evaluation items with superficial information. 
%
The correlation between LLM outputs and human ratings for such an evaluation is high~\cite{murugadoss2025evaluating}, while they often produce holistic evaluations due to the lack of detailed guidelines~\cite{zhou2024llm}. 
Consequently, the evaluation results of LLMs may differ from those of human SOC practitioners.
%
%
The prompt used for high-level evaluation is shown in Fig.~\ref{fig:prompt_all}. 
It consists of three components, i.e., the target analysis report, specific items in the Analyst-wise Checklist, and the evaluation setup, including the task and its method.
The evaluation is performed using a five-point Likert scale, which can provide human-like judgments~\cite{murugadoss2025evaluating} as well as the reliability of a rating scheme~\cite{kusmaryono2022number}.
We note that, within the prompt, only a minimal evaluation method is provided by following the existing prompt structures~\cite{fu2023gptscore}.

\textbf{In-depth Evaluation LLM:}
This enables an LLM to understand important descriptions.
This means that the LLM evaluates analysis reports by approximating the domain-specific contexts as SOC practitioners using the Granularization Guideline.
The In-depth Evaluation LLM offers several advantages. 
First, by interpreting the evaluation process correctly, the LLM can extract rich context from the prompt and easily justify its output.
Second, it can automatically granularize each item using LLMs~\cite{liu2023g-eval,liu2024hd}. 
Nevertheless, we note that evaluations using general LLMs may differ from human judgment due to domain-specific contexts~\cite{chiang2023closer} and prompts whose detailed instructions are sometimes ineffective~\cite{murugadoss2025evaluating}.
To avoid such a situation, the prompt used for the In-depth Evaluation LLM consists of three components: an analysis report to be evaluated, the Analyst-wise Checklist, and the Granularization Guideline, as shown in Fig.~\ref{fig:guide_prompt_all} and ~\ref{fig:prompt_all}. 
The task of this LLM is to granularize each item of the checklist in accordance with the given analysis report. 

\textbf{Integration:}
This step integrates the high-level and in-depth evaluations to evaluate analysis reports. 
Since each evaluation has distinct characteristics, integrating their outputs yields more reliable and interpretable results.
It can also mitigate potential blind spots and biased judgments arising from a single viewpoint by virtue of cross-referencing outputs from different levels of evaluation\cite{liu2024hd}.
The prompt is structured as follows. 
First, the LLM is instructed to generate an initial score for each item in the checklist granularized by the In-depth Evaluation LLM. 
Next, the LLM is given the results of the High-level Evaluation LLM, and then reflects both results to reconsider the initial scores. 
The LLM then outputs a final score with a five-point Likert scale and feedback for the given report. 
The prompt example is shown in Fig.~\ref{fig:prompt_all}. 

\section{Quantitative Evaluation of Analysis Reports through Rating}
\label{sec:quantitativeevaluation}

In this section, we conduct quantitative evaluations to confirm if MESSALA can evaluate analysis reports consistently with veteran SOC practitioners under the evaluation criteria described above, thereby addressing RQ2. 
Specifically, we compare five-point Likert scales evaluated by MESSALA with those by veteran SOC practitioners for analysis reports. 
We first outline the experimental setting, and then present the results.

\subsection{Experimental Setting} \label{sec:exp_setting}
We describe datasets, evaluation metrics, and baselines, including their implementations.

\subsubsection{Datasets} \label{sec:dataset}

We used the following two datasets for this experiments.

\paragraph{(1) Real-world Analysis Reports}

The first dataset is a private dataset as a collection of analysis reports produced by real-world SOCs. 
These reports were gathered from multiple SOCs monitoring three distinct environments, including factories, buildings, and IT infrastructure. 
Each SOC has independent operation and environment and differs substantially in its monitoring scope, IDS configuration, and alert types. 
All the analysis reports contain alerts triggered by network monitoring.
We omit their details due to ethical reasons as described in Section~\ref{sec:ethical}. 

The dataset consists of 40 reports collected between April 2022 and April 2024, where 12 are for factories, 18 are for buildings, and 10 are for IT infrastructure. 
Report formats are independently standardized within each environment, and all the reports conclude that the alert represents a benign case; however, each report describes either events with non-negligible risk or ambiguous behavior, which require confirmation of the respective clients.
Their contents include basic metadata (e.g., date and alert information), an event summary, impact assessment, detailed analysis, items requiring confirmation and action, and recurrence prevention. 
The average length of the reports is about 2,200 characters, and most reports are in PDF format. 
Sensitive information, such as IP addresses and proper names, was anonymized for the use of LLMs in a local environment. 
We do not plan to release the reports themselves due to ethical reason.

\paragraph{(2) Pseudo-Reports}
The second dataset is a pseudo–report dataset generated by an LLM. 
We use this dataset because, at writing this paper, there is no public dataset of analysis reports. 
To guarantee the reproducibility of this evaluation, we adopt an LLM-based pseudo–data generation approach~\cite{pmlr-v252-hao24a}, which is commonly used in domains where real-world data are difficult to release.
In particular, we generated pseudo-reports using MITRE ATT\&CK~\footnote{\url{https://attack.mitre.org/}} patterns. 
Through validating these pseudo-reports, ten reports were selected. 
For pseudo–report generation, we first prepared a report skeleton and a chain-of-thought (CoT) style prompt. 
The skeleton defined a basic structure in which environmental context and alert overview, analysis status, analytical conclusion, recommended actions, and supporting evidence are described in a consistent order. The CoT prompt specified a high-level generation procedure to ensure that these elements are described coherently.
Because the skeleton and procedure alone would result in overly abstract content, we injected external knowledge and experiential information at the early stage of generation to increase concreteness. 

Concretely, we provided the LLM with (i) attack patterns from MITRE ATT\&CK, (ii) typical patterns of detection and response extracted from past analysis reports, and (iii) knowledge about hypothetical critical assets. In addition, for each target attack technique, we had the LLM refer to external reports so that the generated analysis reports would reflect interpretations specific to environments and situations. These steps were implemented by combining prompt chaining with a RAG-style retrieval process.
After generating a large number of pseudo-reports through this pipeline, we used an LLM to automatically score quality and iteratively discarded low-quality samples. Finally, we conducted a manual quality review and selected 10 pseudo-reports, which we used as the second dataset in our evaluation experiments.
Samples of the generation prompts and pseudo-reports are provided in Fig.~\ref{fig:pseudo_report_generation}.

\begin{figure*}[t]	
	\centering
	\includegraphics[width=\linewidth]
    {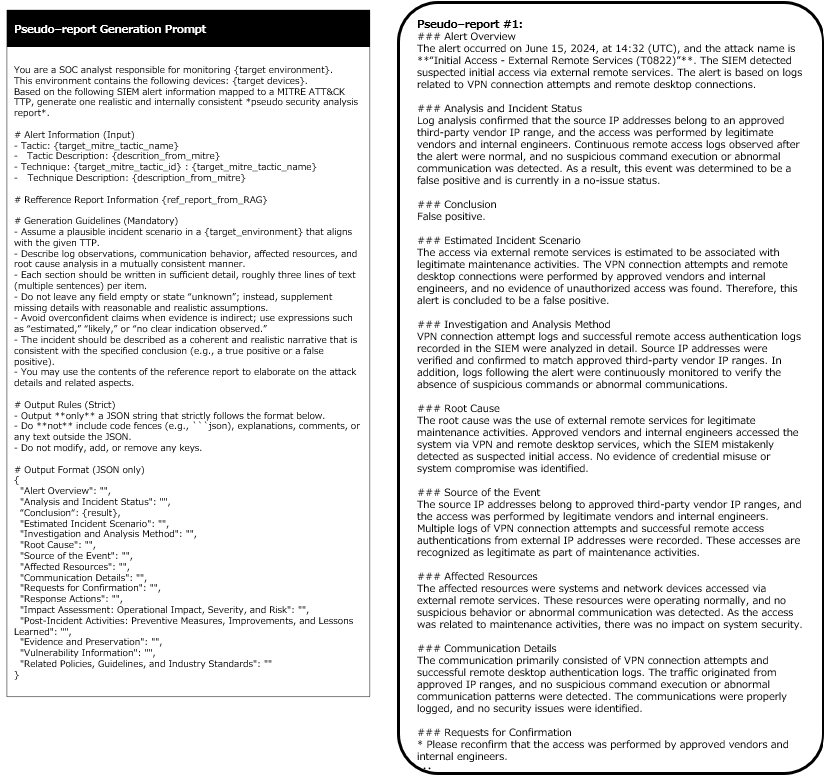}
  \caption{Illustration of pseudo analysis report generation. The left image shows an example prompt used to generate a pseudo analysis report, while the right image presents the resulting pseudo analysis report produced by the LLM based on the prompt.}
  \label{fig:pseudo_report_generation}
\end{figure*}

\paragraph{Human Gold Evaluation of Reports Using the Analyst-wise Checklist}
Using the Analyst-wise Checklist in Section~\ref{sec:EvaluationwithAnalyst-Wise}, we were also able to obtain the evaluation results from the five participants in the user study in Section~\ref{sec:EvaluationwithAnalyst-Wise}. 
Each item in the Analyst-wise Checklist was rated on a five-point Likert scale, and the final score was computed as the average of their individual scores~\cite{chiang2023closer,chiang2023can,gao2023human}.
To align our understanding of the evaluation criteria, the first twenty reports were jointly reviewed. The remaining thirty reports were then processed by the two authors.
For this human gold evaluation, 10 to 15 checklist items were selected, focusing on important contexts of the analysis reports. 
This limitation in the number of items is based on two reasons:
1) Evaluating too many items would increase the workload on the participants and may cause low accuracy; 
2) In realistic operational settings, it is more practical to concentrate on key items aligned with the content of the analysis reports in order to provide meaningful feedback.

In constructing the expert evaluation dataset, we followed three important precautions: 1) To minimize bias, evaluators were assigned only to reports from SOC environments they were not directly responsible for. As a result, each report was typically scored by approximately three evaluators, and their average was used as the final expert rating. 2) To ensure consistency among evaluators, we conducted preliminary briefings on the checklist items and held a Q\&A session using Microsoft Teams to align understanding. 
3) To preserve independence during the evaluation process, participants were not allowed to view each other scores until all evaluations were completed.

\subsubsection{Evaluation Metrics}

Following the discussion in Section~\ref{sec:highlevelevaluation} and a typical setting in LLM-as-a-judge~\cite{liu2024hd,liu2023g-eval,fu2023gptscore}, we measure a five-point Likert scale and compute its correlation coefficients between an LLM and human gold evaluation as the primary evaluation metric. 
We compute the Spearman’s rank correlation~$\rho$, the Kendall’s tau~$\tau$, and the Pearson’s correlation~$r$ as correlation coefficients, as well as the root mean square error (RMSE) to measure deviation in scores.
While prior work~\cite{fu2023gptscore} computes and averages correlations for each individual document, we aggregate all analysis reports and then compute the correlations because the number of checklist items is limited and distinct for each report. 
We use a total of 755 items derived from fifty analysis reports.

\subsubsection{Baseline}
We implemented MESSALA and the following four methods as baselines including also ablation study, i.e., 
Method~1 to Method~4. 
Hyperparameters are temperature $= 0$, and top-$p = 0$.
All methods use the Analyst-wise Checklist.

\textbf{Method~1: Only High-level Evaluation.}
This method performs the high-level evaluation described in Section~\ref{sec:highlevelevaluation}.
It follows GPTScore~\cite{fu2023gptscore} with a typical prompt that simply evaluates the given texts. 
It also utilizes the Analyst-wise Checklist but not the Granularization Guideline. 

\textbf{Method~2: Only In-depth Evaluation without Granularization Guideline.}
This method follows G-Eval~\cite{liu2023g-eval} by granularizing each item in the Analyst-wise Checklist into detailed evaluation criteria. 
It does not use Granularization Guideline.

\textbf{Method~3: Only In-depth Evaluation with Granularization Guideline.}
This method performs the in-depth evaluation using both the Analyst-wise Checklist and the Granularization Guideline compared with the previous methods. 
Note that it incorporates neither the high-level evaluation nor the multi-perspective evaluation.

\textbf{Method~4: Multi-perspective Evaluation without  Granularization Guideline.}
This method integrates Method~1 and Method~2 without Granularization Guideline.

\subsection{Results}


To answer RQ2, we compare MESSALA’s numerical scores with those of the baseline methods. The results are summarized in Table~\ref{tab:experimentresults} and Fig.~\ref{fig:methodscores}.
Overall, MESSALA outperforms the baseline methods in almost all settings, with only a few exceptions (RMSE on gpt‑4o and qwen3 (14B), $\tau$ on gpt‑4.1 mini, and $r$ on gemma3 (12B)), and achieves the best performance across all evaluation metrics when averaged over models.
Consistently, Fig.~\ref{fig:methodscores} shows that the score distribution produced by MESSALA most closely matches that of the human gold evaluations.

These results demonstrate the effectiveness of the two core components of MESSALA: (a) the concretization of evaluation based on the Analyst-wise Checklist and Granularization Guideline, and (b) the integration achieved through multi-perspective evaluation. In particular, integrating high-level evaluation with in-depth evaluation enables judgments to be grounded in concrete descriptions in the reports, rather than relying on superficial characteristics.
The ablation study further clarifies the role of each component.

\textbf{1) Limitations of single-perspective evaluation:}
Method~1, which relies solely on high-level evaluation, depends on coarse-grained information and therefore fails to adequately assess detailed report content. As a result, it tends to assign uniformly higher scores overall (Method~1 in Fig.~\ref{fig:methodscores}).
Methods~2 and 3, which employ only in-depth evaluation, enable more detailed assessment but exhibit distinct failure modes.
In Method~2, the absence of the Granularization Guideline leads to insufficient specification of evaluation targets. This results in vague and generic judgments, and in some cases inappropriately low scores based on perspectives that are not directly relevant to the report content (Method~2 in Fig.~\ref{fig:methodscores}).
In contrast, while Method~3 employs appropriate evaluation perspectives, its excessive level of detail often leads to overly harsh judgments, even for minor issues. This tendency is consistent with the well-known problem of overly detailed prompts reported in prior work~\cite{kim2023prometheus} (Method~3 in Fig.~\ref{fig:methodscores}).

\textbf{2) Effectiveness of multi-perspective evaluation and the importance of the Granularization Guideline:}
A comparison between Method~4 and MESSALA highlights the effectiveness of multi-perspective evaluation itself. As summarized in Table~\ref{tab:experimentresults}, both approaches produce results that are close to human gold evaluations across multiple models, indicating that integrating multiple perspectives is beneficial. However, with a few exceptions, MESSALA consistently outperforms Method~4 in Table~\ref{tab:experimentresults}, suggesting that simple aggregation of multiple evaluation results is not sufficient.
These findings emphasize that, to integrate high-level and in-depth evaluations in a stable and practical manner, it is essential to explicitly connect the two through the Granularization Guideline.
Outputs examples of the evaluations are shown in Fig.~\ref{fig:outputsample}. 

Cost and Runtime Analysis: 
Using gpt-4.1 mini, we evaluated a representative report of 1.6K tokens from our experiments against 10 checklist items, consuming 58.9K input tokens and 11.6K output tokens.
This incurred \$0.042 per report (as of March 2026) with a total runtime of 196 seconds. 
Given that each monitoring environment in our in-house SOC produces only one to two reports requiring review per day and manager feedback averages more than 30 minutes, the cost and runtime are practical.

\begin{figure*}[t]	
	\centering
	\includegraphics[width=\linewidth]{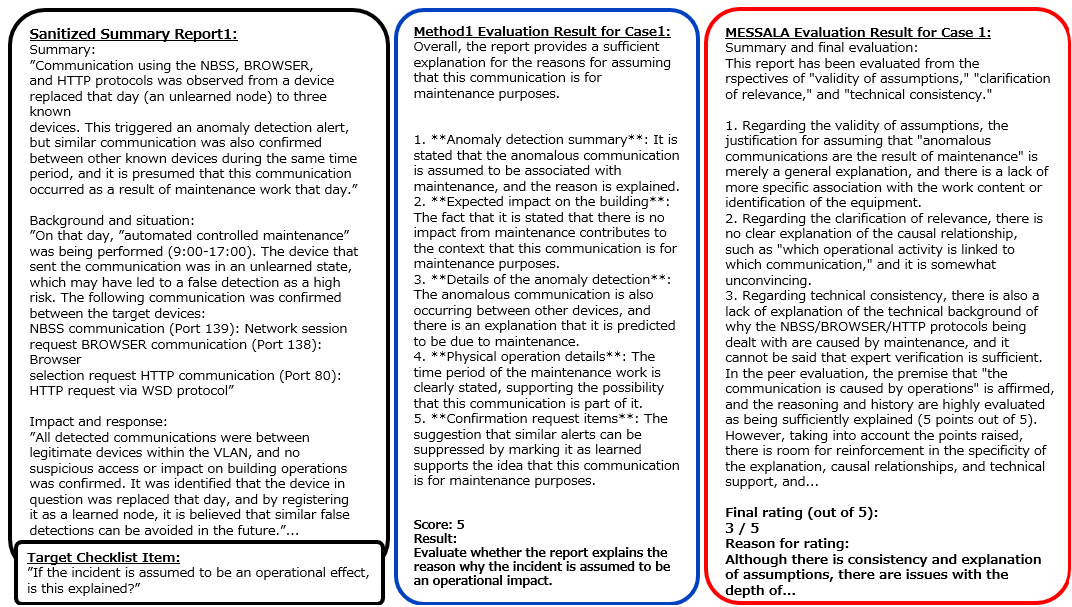}
  \captionof{figure}{LLM-generated feedback examples per method with sanitized reports and checklist items.}
	\label{fig:outputsample}
\end{figure*}

\begin{table*}[t]
  \caption{Results in quantitative evaluations for Open-models (left) and Closed-models (right). 
  The best scores are with bold font, where a smaller score is better for only RMSE. }
    \vspace{-1.5pt}
  \label{tab:experimentresults}
  \centering
  \footnotesize 
  \setlength{\tabcolsep}{4pt}
  \renewcommand{\arraystretch}{1.15}

  \begin{minipage}[t]{0.48\textwidth}
    \centering
  \footnotesize 
    
    
    \vspace{2mm}
    \begin{tabular}{l l c c c c}
      \hline
      \textbf{Model} & \textbf{ } & $\rho$ & $\tau$ & $r$ & RMSE \\
      \hline
      \multirow{5}{*}{gpt-4o}
        & Method~1 & 0.54 & 0.46 & 0.54 & 1.16 \\
        & Method~2  & 0.51 & 0.42 & 0.51 & 1.05 \\
        & Method~3                      & 0.49 & 0.40 & 0.49 & 1.27 \\
        & Method~4                      & 0.55 & 0.45 & 0.53 & \underline{\textbf{1.01}} \\
        & \underline{\textbf{MESSALA}}  & \underline{\textbf{0.60}} & \underline{\textbf{0.49}}
                                        & \underline{\textbf{0.59}} & 1.03 \\
      \hline

      \multirow{5}{*}{gpt-4.1}
        & Method~1 & 0.63 & \underline{\textbf{0.53}} & 0.61 & 1.27 \\
        & Method~2 & 0.60 & 0.50 & 0.57 & 1.07 \\
        & Method~3                      & 0.49 & 0.40 & 0.49 & 1.30 \\
        & Method~4                      & 0.60 & 0.50 & 0.60 & 0.97 \\
        & \underline{\textbf{MESSALA}}  & \underline{\textbf{0.64}} & \underline{\textbf{0.53}}
                                        & \underline{\textbf{0.64}} & \underline{\textbf{0.88}} \\
      \hline

      \multirow{5}{*}{\shortstack{gpt-4.1\\mini}}
        & Method~1 & 0.65 & \underline{\textbf{0.54}} & 0.64 & 1.24 \\
        & Method~2 & 0.54 & 0.42 & 0.55 & 1.23 \\
        & Method~3                      & 0.65 & 0.51 & 0.63 & 0.91 \\
        & Method~4                      & 0.58 & 0.45 & 0.58 & 1.04 \\
        & \underline{\textbf{MESSALA}}  & \underline{\textbf{0.66}} & 0.53
                                        & \underline{\textbf{0.66}} & \underline{\textbf{0.89}} \\
      \hline
      
    \end{tabular}
  \end{minipage}
  \hfill
  \begin{minipage}[t]{0.48\textwidth}
    \centering
  \footnotesize 
    
    
    \vspace{2mm}
    \begin{tabular}{l l c c c c}
      \hline
      \textbf{Model} & $\rho$ & $\tau$ & $r$ & RMSE \\
      \hline

      \multirow{5}{*}{\shortstack{gpt-oss\\(20B)}}
        & 0.56 & 0.44 & 0.55 & 1.27 \\
        & 0.41   & 0.33   & 0.39   & 1.56   \\
                              & 0.56   & 0.43   & 0.54   & 1.16   \\
                              & 0.46   & 0.36   & 0.44   & 1.34   \\
        & \underline{\textbf{0.59}} & \underline{\textbf{0.46}}
                                        & \underline{\textbf{0.59}} & \underline{\textbf{1.13}} \\
      \hline

      \multirow{5}{*}{\shortstack{qwen3\\(14B)}}
         & 0.56 & 0.44 & 0.55 & 1.01 \\
         & 0.52   & 0.40   & 0.51   & 0.97   \\
                              & 0.50   & 0.39   & 0.49   & 1.00   \\
                              & 0.57   & 0.44   & \underline{\textbf{0.56}}   & \underline{\textbf{0.93}}   \\
                               & \underline{\textbf{0.58}} & \underline{\textbf{0.46}} & \underline{\textbf{0.56}} & 1.02 \\
      \hline

      \multirow{5}{*}{\shortstack{gemma3\\(12B)}}
        & 0.28 & 0.23 & 0.28 & 1.52 \\
        & 0.36   & \underline{\textbf{0.30}}   & \underline{\textbf{0.36}}   & 1.12   \\
        & 0.22   & 0.17   & 0.23   & 1.34   \\
        & 0.30   & 0.23   & 0.28   & 1.17   \\
        & \underline{\textbf{0.38}} & \underline{\textbf{0.30}}                & 0.35 & \underline{\textbf{1.06}} \\
      \hline

    \end{tabular}
  \end{minipage}

\end{table*}

\begin{figure}[t]
  \centering
  \begin{subfigure}{0.48\linewidth}
    \centering
    \includegraphics[width=\linewidth]{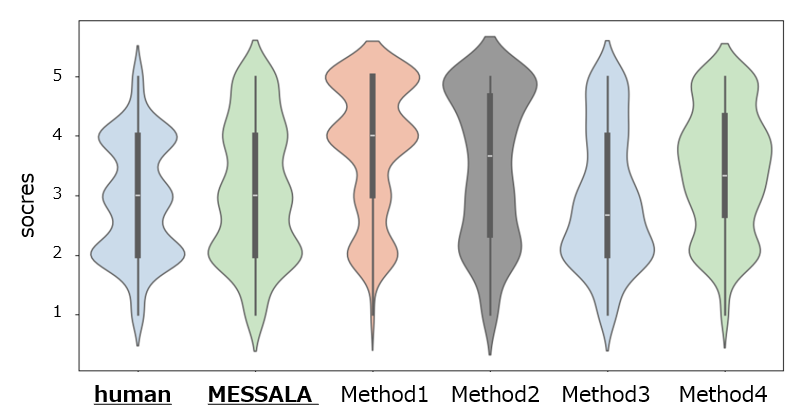}
    \caption{Violin plot of method-wise scores from gpt-4.1.}
    \label{fig:fig1}
  \end{subfigure}
  \hfill
  \begin{subfigure}{0.48\linewidth}
    \centering
    \includegraphics[width=\linewidth]{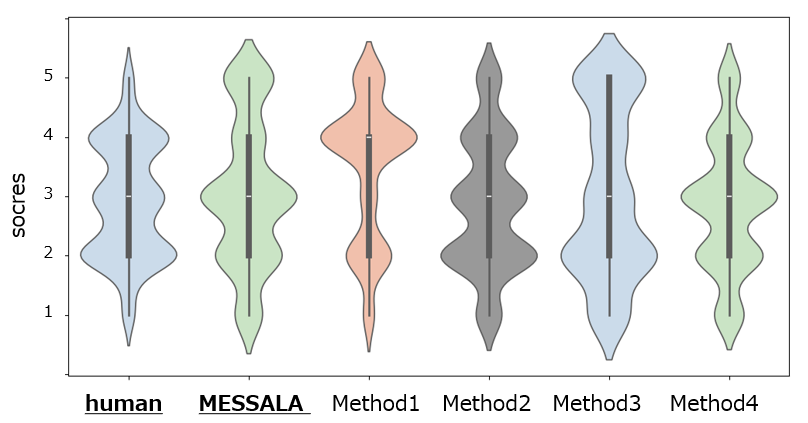}
    \caption{Violin plot of method-wise scores from gpt-4o.}
    \label{fig:fig2}
  \end{subfigure}
  \caption{Violin plots for the distribution of each method. 
  Similar forms of viloins represent similar distributions between each other.}
    \label{fig:methodscores}
\end{figure}

\subsection{Answers to RQ2} \label{sec:answersrq2}
For all the metrics except for a few cases, MESSALA consistently outperforms all the baseline methods: in particular, it consistently achieves a high correlation with the human gold evaluation.
According to prior work~\cite{zhou2024llm,chiang2023closer}, a correlation of approximately 0.6 is regarded as a fairly acceptable level for evaluation results. 
MESSALA achieves correlations of 0.6 on several models, such as gpt-4o and gpt-4.1. 
Consequently, these results indicate that LLMs can rate analysis reports in alignment with human evaluations by virtue of MESSALA, and thus can quantitatively evaluate them from the perspective of SOC practitioners under the criteria as the answer to RQ2. 

\section{Qualitative Evaluation of LLM Feedback Comments}
\label{sec:qualitativeevaluation}

In this section, we conduct experiments for qualitative evaluation in terms of the feedback comments generated by MESSALA. 
The goal of this evaluation is to identify whether the LLM can produce actionable and meaningful feedback from perspective of SOC practitioners. 
This experiment addresses RQ3.
To this end, we design two complementary evaluations. 
The first evaluation examines how the LLM-generated feedback comments to analysis reports are useful for SOC practitioners and LLMs from qualitative standpoints, following prior work~\cite{zubaer2025gpt}. 
The second evaluation examines whether the LLM can correctly point out defects within analysis reports as feedback using a dataset consisting of defect-injected analysis reports. 

\subsection{Multi-metric Rating by Human Experts and an LLM Judge}
We describe Multi-metric Rating evaluation assessed by both human analysts and LLMs as a qualitative evaluation of the feedback comments generated by the LLM.

\subsubsection{Experimental Setting} 

Three participants (UserID~1, 2, and~5), who also took part in the interviews described in Section~\ref{sec:EvaluationwithAnalyst-Wise}, were involved in this evaluation.
They are veteran analysts and managers in SOC of the authors’ affiliation and belong to a different SOC domain, such as factories, buildings, and home appliances. 
This evaluation setting follows that in Section~\ref{sec:quantitativeevaluation} except for the following parts. 
We utilize only Method~1 and MESSALA to reduce the workload of the participants. 
The human gold evaluation uses the Real-world Analysis Report dataset described in Section~\ref{sec:quantitativeevaluation}, whereas Method~1 and MESSALA uses the entire dataset described in Section~\ref{sec:quantitativeevaluation}.
Each participant was assigned six distinct analysis reports, consisting of two reports from each of the three domains, thereby evaluating feedback comments on eighteen analysis reports. 
The task for each report was limited within approximately 10–15 minutes, and the total task per participant was completed around one hour to reduce their workload.
We also applied the same evaluation to an LLM for the entire dataset, which is a common approach for assisting  human evaluation \cite{zubaer2025gpt}. 


Considering the interviews in Section~\ref{sec:EvaluationwithAnalyst-Wise}, we adopt the following metrics:
{\it Overall Usefulness} to measure how feedback is practical; 
{\it Accuracy and Validity} to measure how feedback is precise; 
{\it Specificity and Concreteness} to measure how feedback is actionable; and {\it Support for Novices/Veterans} to measure how feedback supports analysts with different experience levels from novices to veterans. 
We conducted a user questionnaire on the above metrics using a 5-point Likert scale, where ``1" indicates ``very low" and ``5" indicates ``very high". 
Participants can also provide free-text comments to explain their ratings. 
We also utilize Method~1 described in the previous section as a baseline.
User feedback evaluation items are presented in at Table~\ref{tab:feedback_evaluation}.
\begin{table*}[t!]
\centering
\footnotesize
\caption{User Feedback Evaluation Items for LLM-Generated Reports}
\label{tab:feedback_evaluation}
\begin{tabularx}{\textwidth}{|l|l|X|}
\hline
\textbf{Category} & \textbf{No.} & \textbf{Summary} \\
\hline
Overall Usefulness & F-1. &
Do you find the LLM-generated feedback useful overall for analysts? \\
Accuracy and Validity & F-2. &
Does the feedback accurately identify errors or missing information in the report? How well does it align with expert-level feedback? \\
Specificity and Concreteness & F-3. &
Does the evaluation provide concrete improvement points and reproducible advice? \\
Support for Novices & F-4. &
Is the feedback understandable and helpful even for novice analysts? \\
Support for Veterans & F-5. &
Is the feedback still useful for highly veteran analysts? \\
\hline
\end{tabularx}
\end{table*}

\paragraph{Feedback comments by LLM}
We limit the length of feedback comments to 500 tokens by summarizing the results derived from each method.
In this study, since multiple checklist items are evaluated for a single report, presenting all evaluation results as-is would result in an excessive amount of information.
Accordingly, we chose to follow the appropriate amount of feedback indicated in prior studies \cite{zhou2024llm,liang2024can}.
The summarized feedback comments are created as critical comments, focusing on points that received particularly low evaluation scores.
Although best practices~\cite{hattie2007power} for feedback generally recommend including positive comments to support user motivation, analysis reports often require rapid actions. 
Therefore, feedback comments in this evaluation are restricted to checklist items that are less than or equal to ``3". The following instructions are given in the prompt: 
“Based on the following evaluation results, generate feedback comments for improving the report in no more than 500 tokens. Preserve as much information from the original evaluation as possible, describe the issues concretely, and do not add any new information.”

\subsubsection{Results}

\begin{table}[t!]
    \centering
    \caption{Human and LLM ratings (1–5) for LLM-based evaluation results.}
    \label{tab:fbuserevaluation}
    \begin{tabular}{lcccc}
        \hline
        \multirow{2}{*}{\textbf{User Feedback Aspect}} 
            & \multicolumn{2}{c}{\textbf{Human Evaluation}} 
            & \multicolumn{2}{c}{\textbf{LLM Evaluation}} \\
        & \textbf{Method~1} & \textbf{MESSALA} 
        & \textbf{Method~1} & \textbf{MESSALA} \\
        \hline
        Overall Usefulness 
            & 3.00 & \textbf{4.00} & 4.03 & \textbf{4.35} \\
        Accuracy and Validity 
            & \textbf{4.00} & 3.67 & 4.15 & \textbf{4.41} \\
        Specificity and Concreteness 
            & 2.67 & \textbf{4.67} & 4.03 & \textbf{4.51} \\
        Support for Novices 
            & 4.00 & \textbf{4.67} & 3.48 & \textbf{3.65} \\
        Support for Veterans 
            & 3.00 & \textbf{3.67} & 4.08 & \textbf{4.35} \\
        \hline
    \end{tabular}
\end{table}

\begin{figure}[t]	
	\centering
  \includegraphics[width=0.5\linewidth]{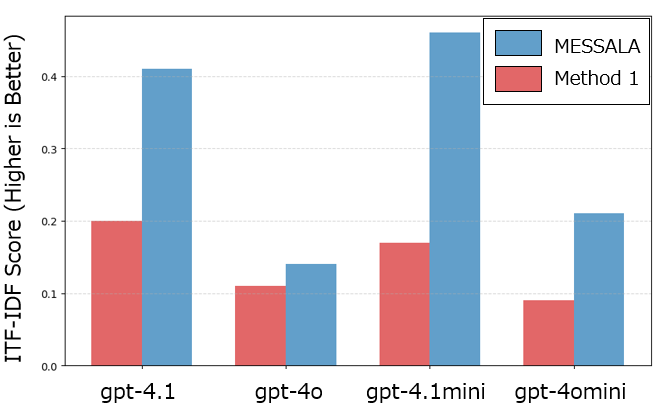}
  \caption{Differences in comment content across reports measured by ITF-IDF.}
  \label{fig:itf-idf}
\end{figure}

The results are shown in Table~\ref{tab:fbuserevaluation}. 
Overall, MESSALA obtains higher ratings, except for ``Accuracy and Validity". 
We describe the reason for each metric below.

\textbf{Overall Usefulness: }
MESSALA was rated more useful than Method~1 because it more clearly pointed out concrete improvement points in analysis reports. 
As one participant noted, \emph{``Compared to Method~1, evaluation and suggestions for improvement by MESSALA are more specific and easier to understand.''} -- UserID~1.
In contrast, the feedback by Method~1 was readable but often inconsistent and generic: \emph{``The report is concise and approachable, but its feedback lacks consistency compared to MESSALA, reducing its usefulness.''} -- UserID~5.

\textbf{Accuracy and Validity: }
Method~1 received a slightly higher score with 4.0 than MESSALA with 3.67. Although its feedback remained superficial, it was generally reasonable with few errors or contradictions, which contributed to favorable judgments.
Whereas MESSALA is more detailed, it sometimes focuses on issues that the participants consider minor, leading to slightly lower scores: \emph{``MESSALA sometimes overemphasizes minor issues, resulting in overly detailed feedback on less important points.''} -- UserID~5.
Overall, MESSALA can provide detailed feedback but still struggles to prioritize the most critical issues compared with the human gold evaluation. 
More established guidelines for each checklist item may improve this point.

\textbf{Specificity and Concreteness:}
MESSALA outperformed Method~1 on this metric. 
The participants evaluated how the feedback comments aligned each key point with its suggestion, making LLM outputs easier to translate into actionable items: \emph{``In MESSALA, key points and suggested improvements are mapped one-to-one, making action items easy to derive.''} -- UserID~5.
Method~1 was reasonable but often lacked explicit links between evaluations and feedback, and its wording was sometimes vague. 
The participants preferred more direct expressions such as ``missing,'' ``unclear,'' or ``insufficient'': \emph{``More direct feedback -- such as `missing,' `unclear,' or `insufficient' -- is necessary rather than vague comments.''} -- UserID~2.
We also measured the Specificity and Concreteness of the feedback comments using the ITF-IDF metric~\cite{du2024llms} to assess whether detailed and specific comments are obtained. 
It enables us to evaluate whether similar comments are repeatedly generated across different reports.
The results are shown in Fig.~\ref{fig:itf-idf}. 
According to the figure, we observe that MESSALA produces report-specific comments, which are similar to the human gold evaluation.

\textbf{Support for Novices and Veterans:}
MESSALA was also rated more than Method~1 as supporting both novice and veteran SOC practitioners. 
Its feedback is structured for key points, and hence novice SOC practitioners can quickly identify which parts of a report should be revised: \emph{``Because MESSALA feedback is focused and concise, it's more accessible to novices than Method~1, which requires reading the full text to understand.''} -- USER\_ID\_2.
For veteran SOC practitioners, one-by-one feedback from MESSALA makes the resulting feedback easier to extract relevant information than Method~1. 
However, the participant noted that both methods tend to misinterpret technical terms and rely on indirect expressions, highlighting the need for stronger domain knowledge in LLMs:
\emph{``Both methods sometimes independently interpret technical terms, which obscures their intended meaning, and although they extract implicit information, their feedback is less direct and targeted than that of human experts when it comes to improving report quality.''} — UserID~5

\subsection{Evaluation on Defect-injected Analysis Reports}
This section conducts evaluation on defect-injected analysis reports to examine how each method identifies and explains defects and their reasons, following prior works~\cite{liu2023reviewergpt,d2024marg}.
For effective evaluations of analysis reports, a method must accurately identify specific points for improvement in order to facilitate report refinement. 
Prior work~\cite{liang2024can} has pointed out a limitation that evaluations by LLMs are sometimes overly lenient and may fail to highlight critical defects. 
Our goal is thus to investigate whether MESSALA can mitigate this limitation and provide more actionable and meaningful feedback.

\subsubsection{Defect-injected Analysis Reports}

The defect-injected analysis report dataset is designed to evaluate whether an LLM can accurately identify quality defects in SOC analysis reports. The dataset consists of reports with realistic, context-dependent defects commonly observed in SOC environments, such as missing contextual information, insufficient justification for judgments, flawed causal reasoning, and unclear or impractical response descriptions. Formally, we define a defect as a case where at least one of the eleven evaluation perspectives derived in Section~\ref{sec:EvaluationwithAnalyst-Wise} is missing or unsatisfied.


Based on defect categories proposed in prior work~\cite{liu2023reviewergpt,d2024marg} and reference review comments collected from SOCs within the authors’ organization, we consolidated the original eleven evaluation perspectives into four defect categories, with confirmation by SOC practitioners. Table~\ref{tab:defect_categories_short} summarizes the relationship between the four defect categories and the original evaluation perspectives.

\begin{table*}[tb]
  \centering
  \caption{Defect categories and their corresponding practitioner-oriented evaluation perspectives}
  \label{tab:defect_categories_short}
  \small
  \setlength{\tabcolsep}{6pt}
  \begin{tabularx}{\textwidth}{p{0.23\textwidth} X p{0.22\textwidth}}
    \toprule
    \textbf{Defect Category} 
      & \textbf{Short Description} 
      & \textbf{Related Evaluation Perspectives} \\
    \midrule

    Opaque Decision Rationale
      & Key judgments and conclusions lack sufficient justification or coherent reasoning, undermining confidence in the final assessment.
      & (1), (10), (4) \\

    Unverifiable or One-Sided Analysis
      & Analytical claims are made without sufficient verification, such as baselines, comparisons, or alternative explanations.
      & (8), (10), (4) \\

    Context-Agnostic Technical Interpretation
      & Technical observations are interpreted without adequate consideration of on-site context, such as asset roles, operational background, or constraints.
      & (5), (6), (7), (9), (4) \\

    Non-Actionable Outcome Presentation
      & Analysis results fail to translate into concrete, prioritized actions or clearly articulated impact and escalation implications.
      & (2), (3), (11), (4) \\

    \bottomrule
  \end{tabularx}
\end{table*}

Following these categories, we inject one or more defects into high-quality analysis reports and define reference review comments that explicitly point out the inserted defects. These reference review comments are derived from real SOC analysis reports, for example: “The report does not explain why the communication occurred, nor does it assess whether the communication or the involved endpoint is anomalous.” To construct defect-injected reports, we first selected a subset of analysis reports that were evaluated as high quality by both human analysts and an LLM. Each report was then modified according to predefined injection rules, while preserving the overall structure and writing style by only altering the relevant portions.

These datasets were constructed with the assistance of LLMs, whose outputs were subsequently reviewed and refined by the authors, and some defect samples were further validated through reviews by SOC practitioners.
We finally utilized 44 analysis reports as the dataset. 
Furthermore, we conduct a preliminary validation to verify that quality deterioration of analysis reports caused by the injected defects is measurable. 
For each defect category, we compare evaluation scores between clean and defect-injected analysis reports derived from the same original report, and then identify whether score deteriorates in the corresponding evaluation perspectives while scores on unrelated evaluation perspectives remain unchanged. 
Defect injection and validation are conducted independently and the LLM performs scoring without access to defect labels or human verification results.
The results are presented in Table~\ref{tab:precheck_avgdrop}.

Importantly, defect categories do not correspond to checklist items in a one-to-one manner. A single defect often affects multiple quality dimensions simultaneously, such as justification, contextualization, and decision support. Therefore, we intentionally avoid defining strict mappings between defect categories and checklist items. Although the defects are artificially injected, they are designed to be subtle and compound, reflecting the ambiguous and intertwined nature of quality issues in real SOC analysis reports. This dataset focuses on evaluating the ability to identify and assess such defects, which is a prerequisite for generating meaningful improvement suggestions. An example of a defect-injected report is shown in Fig.~\ref{fig:defect_injection_compare}.

\begin{table}[t]
  \centering
  \caption{Sample output of preliminary validation showing average score changes}
  \label{tab:precheck_avgdrop}
  \small
  \setlength{\tabcolsep}{5pt}
  \begin{tabular}{p{6cm} cc}
    \toprule
    \textbf{Defect Category} 
      & \makecell{\textbf{Avg. Score Drop}\\\textbf{(Targeted)}} 
      & \makecell{\textbf{Avg. Score Drop}\\\textbf{(Unrelated)}} \\
    \midrule
    Opaque Decision Rationale 
      & $-1.1$ & $-0.1$ \\
    Unverifiable or One-Sided Analysis 
      & $-0.9$ & $-0.2$ \\
    Context-Agnostic Technical Interpretation 
      & $-1.3$ & $-0.1$ \\
    Non-Actionable Outcome Presentation 
      & $-1.0$ & $-0.1$ \\
    \bottomrule
  \end{tabular}
\end{table}

\begin{figure*}[t]	
	\centering
\includegraphics[width=\linewidth]{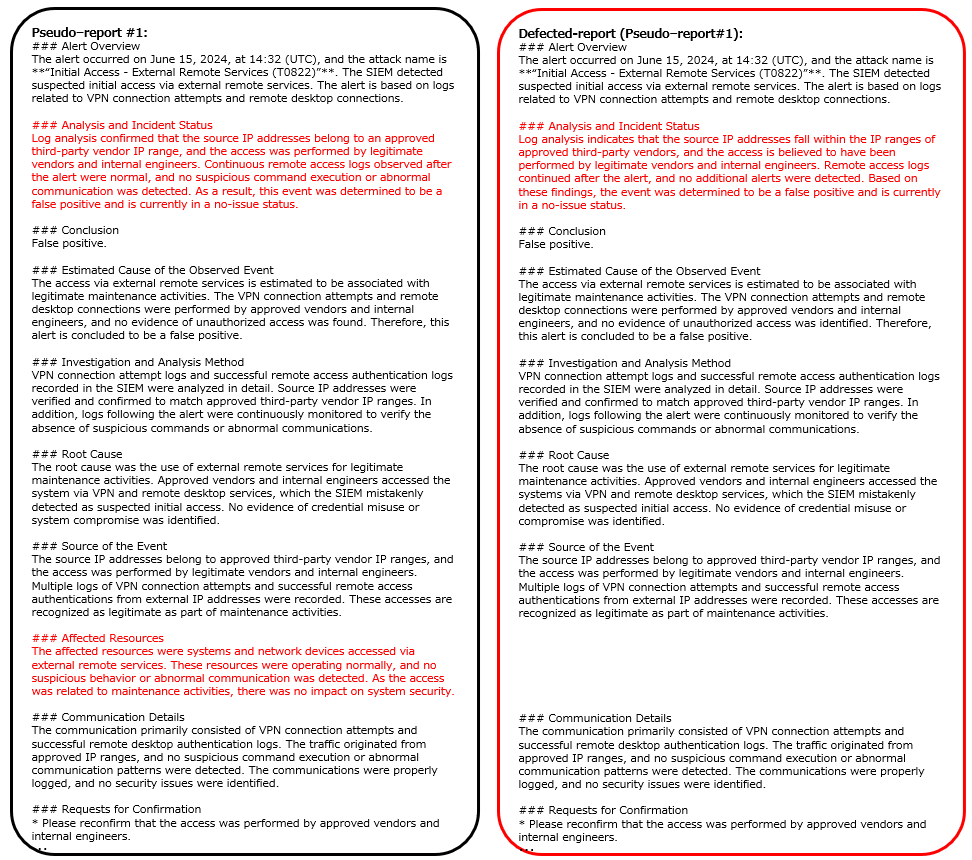}
	\caption{Comparison of the report excerpts before and after defect injection. Red text highlights the injected differences.}
\label{fig:defect_injection_compare}
\end{figure*}

\paragraph{Defect-injection Rule}
In the following, we present an example of defect-injection rules, focusing on the category: \emph{Opaque Decision Rationale}. Similar rules are defined for the other defect categories, and defects are primarily injected into the reports through manual modification.

Rule 1) Preserve the final conclusions while obscuring the decision rationale: 
The original analytical conclusions, such as the presence or absence of an attack, the assessed risk level, and the final judgment outcome, are kept unchanged.
Defects are injected exclusively by degrading the explanation of how these conclusions are derived, rather than altering the conclusions themselves.

Rule 2) Remove explicit assumptions, baselines, and evaluation criteria supporting the judgment:
The assumptions, comparison baselines, and evaluation criteria that underpin the analyst’s judgment are intentionally removed or abstracted.
As a result, the framework used to assess the situation is no longer explicitly stated.

Rule 3) Omit the logical linkage between observed evidence and conclusions:
Observed facts, such as logs, communication behaviors, or events, are preserved in the report.
However, explicit explanations describing how these observations lead to the stated conclusions are omitted, creating gaps in the reasoning process.

Rule 4) Weaken or abstract causal and inferential expressions:
Explicit causal or inferential connectives (e.g., because, therefore, based on, due to, for this reason) are removed or replaced with vague and non-specific expressions, thereby obscuring the reasoning flow.

Rule 5) Present conclusions in a definitive manner without sufficient justification:
Judgments are expressed in confident and assertive terms, while providing little or no supporting rationale.
This results in reports that appear conclusive but lack sufficient justification for reviewers to assess the validity or soundness of the analysis.

\subsubsection{Experimental Setting} 
We use the defect-injected analysis report dataset described above to examine how each method identifies defects in analysis reports. 
We focus on whether it can generate feedback beyond superficial detection of defects and capture issues that should be improved according to the reference review comments.

The evaluation targets the feedback comments generated by the LLM for each defect-injected analysis report.
We assess, through manual inspection, that these comments capture the reference review comments. 
Specifically, each feedback comment is categorized into three levels: (1) clearly covered, where it sufficiently captures the reference review comment; (2) partially covered, where it partially captures an essential part of the reference review comments despite differences in focus and wording; and (3) not covered, where it captures nothing of the reference review comment. 
The criteria for determining whether a comment covers defects are agreed upon by the authors in advance.
As a baseline, we use Method~1 described in Section~\ref{sec:quantitativeevaluation}. 
Comparison with this baseline enables us to assess how MESSALA generates more meaningful feedback. 

\begin{table}[t]
  \centering
  \caption{Coverage results (left) and defect detection breakdown (right).}
  \label{tab:coveragedefects}

  \begin{minipage}[t]{0.47\linewidth}
    \centering
    \caption*{(a) Coverage of Reference Review Comments}
    \begin{tabular}{lccc}
      \toprule
      Method & Clearly & Partially & Not \\
      \midrule
      \textbf{MESSALA} & \textbf{\underline{29}} & 9 & \textbf{\underline{6}} \\
      Method1 & 16 & 15 & 13 \\
      \bottomrule
    \end{tabular}
  \end{minipage}
  \hfill
\begin{minipage}[t]{0.47\linewidth}
  \centering
  \caption*{(b) MESSALA: Defect Category Breakdown}
  \begin{tabular}{lcc}
    \toprule
    Category & \begin{tabular}[c]{@{}c@{}}Partially \\/ Clearly\end{tabular} & \begin{tabular}[c]{@{}c@{}}Not \\covered\end{tabular} \\
    \midrule
    Opaque Decision Rationale   & 10 & 2 \\
    Unverifiable / One-Sided    &  7 & 2 \\
    Context-Agnostic            & 12 & 2 \\
    Non-Actionable Outcome      &  9 & 0 \\ 
    \bottomrule
    \label{tab:breakcate}
  \end{tabular}
\end{minipage}

\end{table}

\subsubsection{Results}

Table \ref{tab:coveragedefects} presents the defect identification results for all methods.
MESSALA identifies 86.4\% of the injected defects, exceeding Method~1 across all defect
categories. This performance difference reflects systematic discrepancies in how each
method evaluates the relationship between conclusions, supporting evidence, contextual
information, and decision-relevant actions.

\paragraph{Opaque Decision Rationale}
In this category, we compare methods based on examples that assess whether conclusions are described with sufficient supporting evidence.

Fig.~\ref{fig:comment_box} shows an excerpt of evaluation comments generated by each method on a report that contains the following defect: \emph{``although the report mentions device details and the presence or absence of related vulnerabilities and concludes that the alert is not problematic, it fails to explain why the alert was triggered and lacks consideration of whether the corresponding communication has any operational impact.''}

Method~1 evaluates the report as sufficient for the checklist item related to alert causation, primarily because it mentions the roles of the devices and their associated vulnerabilities.
In contrast, MESSALA assigns low evaluation scores when the rationale supporting the conclusion is insufficient. Based on granularized evaluation items, MESSALA identifies and explicitly points out that the report fails to explain how the detected device anomalies or state changes would impact building operations.

\paragraph{Unverifiable or One-Sided Analysis}
This category compares methods based on whether a plausible analytical process or methodology is clearly applied to the observed event.

Fig.~\ref{fig:comment_box2} shows an excerpt of evaluation comments from each method on a report that contains the following defect: \emph{''although communication potentially corresponding to a vulnerability-exploiting attack is observed, the report provides no detailed analysis and concludes that the communication is not problematic simply because it is used in routine operations. As a result, the analysis omits an assessment of the anomalous nature of the potentially vulnerable communication itself and a careful evaluation of whether it indeed corresponds to legitimate operational traffic.``}

Method~1 is able to partially identify specific defects in the report; however, its feedback is limited to high-level observations and lacks practical guidance. It does not clearly indicate which analytical evidence is missing or how analysts should re-examine the underlying assumptions, resulting in feedback that is insufficient for improving report quality.
Compared with Method~1, MESSALA provides more informative and practical feedback by explicitly identifying insufficient analytical reasoning and inadequately justified false-positive determinations. Rather than accepting high-level operational explanations at face value, MESSALA critically examines the relationship between detected anomalous communication patterns and legitimate business traffic, thereby revealing detailed analytical gaps that Method~1 fails to capture.

\begin{figure}[t]
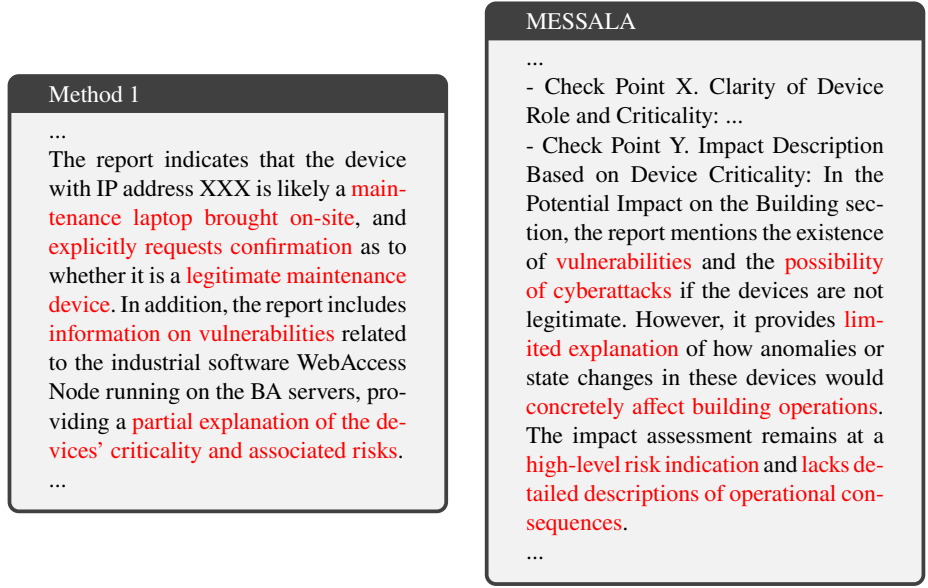

\centering
\caption{Comparison of Evaluation Comments: Opaque Decision Rationale category}
\label{fig:comment_box}
\begin{minipage}{0.48\linewidth}
\begin{tcolorbox}[title=Method~1]
...\\
The report indicates that the device
with IP address XXX is likely a
\textcolor{red}{maintenance laptop brought on-site},
and \textcolor{red}{explicitly requests confirmation}
as to whether it is a
\textcolor{red}{legitimate maintenance device}.
In addition, the report includes
\textcolor{red}{information on vulnerabilities}
related to the industrial software
WebAccess Node running on the BA servers,
providing a
\textcolor{red}{partial explanation of the devices’ criticality and associated risks}.
\\...
\end{tcolorbox}
\end{minipage}
\hfill
\begin{minipage}{0.48\linewidth}
\begin{tcolorbox}[title=MESSALA]
...\\
- Check Point~X. Clarity of Device Role and Criticality: ...\\
- Check Point Y. Impact Description
Based on Device Criticality:
In the Potential Impact on the Building section,
the report mentions the existence of
\textcolor{red}{vulnerabilities}
and the \textcolor{red}{possibility of cyberattacks}
if the devices are not legitimate.
However, it provides
\textcolor{red}{limited explanation}
of how anomalies or state changes in these devices
would \textcolor{red}{concretely affect building operations}.
The impact assessment remains at a
\textcolor{red}{high-level risk indication}
and \textcolor{red}{lacks detailed descriptions of operational consequences}.
\\...
\end{tcolorbox}
\end{minipage}
\end{figure}

\begin{figure}[tb]
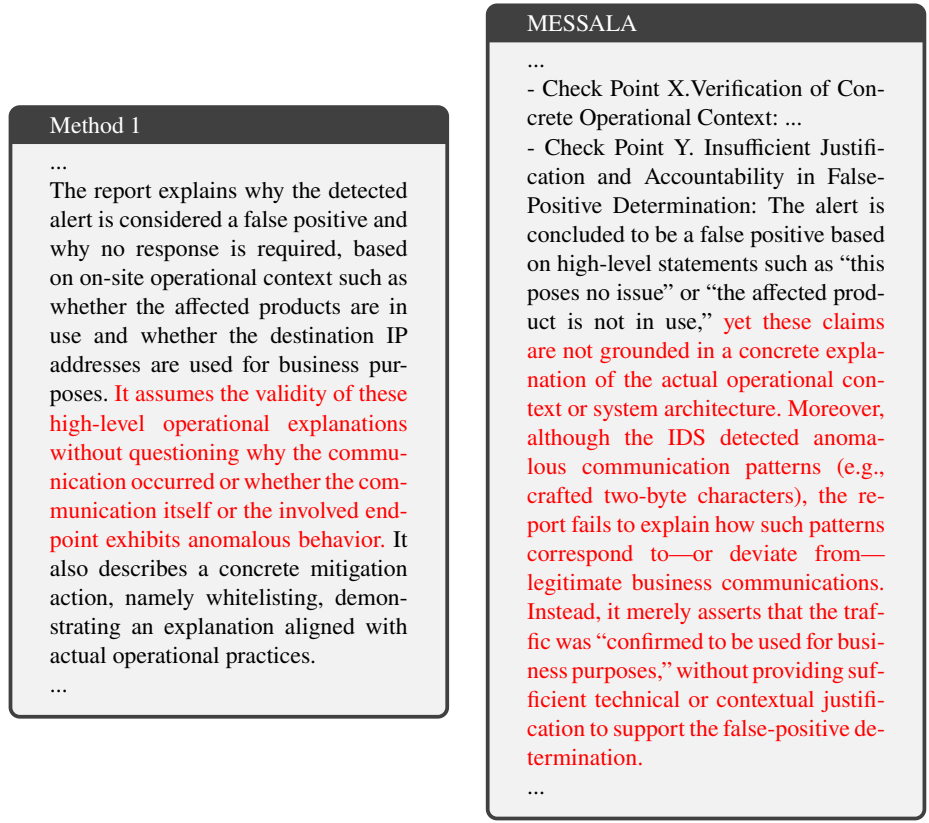

\centering
\caption{Comparison of Evaluation Comments: Unverifiable or One-Sided Analysis category}
\label{fig:comment_box2}
\begin{minipage}{0.48\linewidth}
\begin{tcolorbox}[title=Method~1]
...\\
The report explains why the detected alert is considered a false positive and why no response is required, based on on-site operational context such as whether the affected products are in use and whether the destination IP addresses are used for business purposes.
\textcolor{red}{It assumes the validity of these high-level operational explanations without questioning why the communication occurred or whether the communication itself or the involved endpoint exhibits anomalous behavior.}
It also describes a concrete mitigation action, namely whitelisting, demonstrating an explanation aligned with actual operational practices.
\\...
\end{tcolorbox}
\end{minipage}
\hfill
\begin{minipage}{0.48\linewidth}
\begin{tcolorbox}[title=MESSALA]
...\\
- Check Point~X.Verification of Concrete Operational Context: ...\\
- Check Point~Y. Insufficient Justification and Accountability in False-Positive Determination: 
The alert is concluded to be a false positive based on high-level statements such as ``this poses no issue'' or ``the affected product is not in use,''
\textcolor{red}{yet these claims are not grounded in a concrete explanation of the actual operational context or system architecture.}
\textcolor{red}{Moreover, although the IDS detected anomalous communication patterns (e.g., crafted two-byte characters), the report fails to explain how such patterns correspond to---or deviate from---legitimate business communications.}
\textcolor{red}{Instead, it merely asserts that the traffic was ``confirmed to be used for business purposes,'' without providing sufficient technical or contextual justification to support the false-positive determination.}
\\...
\end{tcolorbox}
\end{minipage}
\end{figure}

Similar evaluation patterns were also observed in the categories of Context-Agnostic Technical Interpretation and Non-Actionable Outcome Presentation.
Method~1 tended to evaluate reports favorably as long as the technical descriptions were accurate or comprehensive, even when the operational context was insufficient or the analysis results were not linked to concrete follow-up actions. Moreover, although issues were sometimes pointed out, the feedback was often high-level and lacked specificity.
In contrast, MESSALA consistently assigns low scores to feedback comments that are technically correct but ignore contextual considerations or fail to provide actionable items for decision making, and delivers more detailed and concrete feedback.

\subsection{Answers to RQ3} \label{sec:answersrq3}
Our results show that MESSALA can generate appropriate feedback comments aligned with veteran SOC practitioners based on the evaluation criteria. 
In the multi-metric rating, feedback comments generated by MESSALA were consistently rated higher by veteran SOC practitioners across all the categories except for Accuracy and Validity. 
It indicates that LLMs can provide actionable feedback for analysis reports compared with the existing method, and thus can quantitatively evaluate them through their feedback comments from the perspective of SOC practitioners by virtue of MESSALA. 

Moreover, the defect-injection evaluation demonstrates that MESSALA can identify defects on analysis itself in addition to providing plausible explanations. While Method~1 tends to assign high scores to reports whose content is consistent but less justified, MESSALA assigns low scores to them by taking the lack of evidence and actionable items into account, thereby correctly identifying 86.4\% of the injected defects.

\endgroup

\begingroup

\section{Limitations and Ethical Consideration}

\subsubsection{Limitations}
Our study has several limitations that need to be addressed in future work.
First, we utilized private datasets as analysis reports. 
While pseudo-reports are also used to support reproducibility, the number of reports and their length are limited, and hence, we need to extend them into a more diverse setting. 
Second, the Analyst-wise Checklist contains a large number of items, and the important evaluation items may differ depending on the reports. Future work will concentrate on a mechanism that automatically selects these items, although they were manually selected in this paper.
Third, the number of SOC practitioners involved in the human gold evaluation was limited. 
To confirm the impact of MESSALA on reducing analysis workload and improving the quality of analysis reports in real-world environments, we also need to conduct further investigation with a larger number of participants.
Finally, the evaluation perspectives in this paper, including the guidelines and defect categories, are based on the literature review and the semi-structured interviews with practitioners and do not fully cover the evaluation processes and defect patterns across all real-world SOCs. 
We are in the process of refining feedback from versatile operation contexts. 

\subsubsection{Ethical Considerations} \label{sec:ethical}

We discuss ethics in interviews and analysis reports below. 

\textbf{Interview:}
Regarding interview data in this paper, we informed the participants in advance of the research objectives, interview procedures, scope of data use, and privacy protection policy, and then obtained their consent. 
The participants were volunteers, and we explained that they could withdraw at any time without any disadvantage under the informed consent. 
Audio recordings of the interviews were used solely in transcription and the qualitative analysis for the informed research objectives and were stored in a local environment accessible only to the authors. 
During the coding process, we removed any information that leads to re-identification, such as personal and organization names including their project details, and anonymized them to pseudonym IDs.

\textbf{Analysis Reports:}
Before using the analysis reports in this paper, we explained the following points, including the use of an LLM service for them, regarding confidentiality and research ethics to stakeholders responsible for each SOC, and then obtained their consent to conduct the experiments. 
In accordance with the principle of data minimization, two of the authors manually anonymized sensitive information, such as proper names, and checked the risk that individual devices and organizations could be re-identified. 
Only the anonymized data was then submitted to an LLM service for which our affiliation has a formal contract (i.e., prohibiting secondary use, third-party provision, and data retention) and obtained the security approvals. 
Access to the data and any use beyond the scope of this paper were prohibited even within the authors’ affiliation, thereby reducing the risk arising from the evaluations described above.

\section{Conclusion} \label{sec:conclusion}

In this paper, we discussed evaluations of analysis reports in SOCs using LLMs. 
We first designed the Analyst-wise Checklist by identifying evaluation criteria for analysis reports in SOCs from multiple perspectives of veteran SOC practitioners as the answer to RQ1 through the literature review and interviews with SOC practitioners. 
Next, we proposed \textit{MESSALA}, which can maximize the evaluation of analysis reports with feedback comments from the perspective of human SOC practitioners. 
We also conducted quantitative evaluations of analysis reports using MESSALA and demonstrated that MESSALA can rate the evaluation criteria for analysis reports, thereby enabling LLMs to quantitatively evaluate the reports compared with the existing methods as the answer to RQ2. 
Finally, we conducted qualitative evaluations of analysis reports using MESSALA in terms of the feedback comments on the reports. 
We then showed that MESSALA can qualitatively evaluate analysis reports and provide actionable feedback comments under the evaluation criteria as the answer to RQ3. 
We plan to conduct further experiments across diverse settings, including larger sample sizes and feedback comments from diverse operation contexts, as well as additional deployments in real-world SOC environments.  
\endgroup

\newtcolorbox{boxA}{
    breakable,
    enhanced,
    colback = white,
    boxrule = 1.5pt,
    colframe = black,
    rounded corners,
    arc = 5pt   
}
\newtcolorbox{boxB}{
    breakable,
    enhanced,
    colback = white,
    boxrule = 1.5pt,
    colframe = red,
    rounded corners,
    arc = 5pt   
}
\newtcolorbox{boxC}{
    breakable,
    enhanced,
    colback = white,
    boxrule = 1.5pt,
    colframe = blue,
    rounded corners,
    arc = 5pt   
}
\newtcolorbox{boxD}{
    breakable,
    enhanced,
    colback = white,
    boxrule = 1.5pt,
    colframe = green,
    rounded corners,
    arc = 5pt   
}

\bibliographystyle{splncs04}
\bibliography{main}

\end{document}